\documentclass[11pt]{article}
\usepackage{amsmath,amssymb,color,graphics,epsfig}

\makeatletter
\@addtoreset{equation}{section}
\makeatother

\usepackage{amsfonts}

\newcommand{\hoch}[1]{$\, ^{#1}$}

\newcommand{\be}{\begin{equation}}
\newcommand{\ee}{\end{equation}}
\newcommand{\bea}{\setlength\arraycolsep{2pt} \begin{eqnarray}}
\newcommand{\eea}{\end{eqnarray}}
\newcommand{\nn}{\nonumber}

\def\im{{{\rm i}}}

\def\ft#1#2{{\textstyle{\frac{\scriptstyle #1}{\scriptstyle #2} } }}
\def\fft#1#2{{\frac{#1}{#2}}}

\def\R{{\mathbb{R}}}
\def\C{{\mathbb C}}

\newcommand{\beq}{\begin{equation}}
\newcommand{\eq}{\end{equation}}
\newcommand{\ea}{\end{eqnarray}}

\def\0{{\sst{(0)}}}
\def\1{{\sst{(1)}}}
\def\2{{\sst{(2)}}}
\def\3{{\sst{(3)}}}
\def\4{{\sst{(4)}}}
\def\5{{\sst{(5)}}}
\def\6{{\sst{(6)}}}
\def\7{{\sst{(7)}}}
\def\8{{\sst{(8)}}}
\def\sst#1{{\scriptscriptstyle #1}}

\def\del{{\partial}}

\begin{document}

\begin{flushright}
\hfill{MIFPA-14-31}
\end{flushright}

\vspace{25pt}
\begin{center}
{\large {\bf Superfluid and Metamagnetic Phase Transitions \\
in $\omega$-deformed Gauged Supergravity}}

\vspace{10pt}
S. Cremonini\hoch{1}, Y. Pang\hoch{1}, C.N. Pope\hoch{1,2} and
J. Rong\hoch{1}

\vspace{10pt}

\hoch{1} {\it George P. \& Cynthia Woods Mitchell  Institute
for Fundamental Physics and Astronomy,\\
Texas A\&M University, College Station, TX 77843, USA}

\vspace{10pt}

\hoch{2}{\it DAMTP, Centre for Mathematical Sciences,
 Cambridge University,\\  Wilberforce Road, Cambridge CB3 OWA, UK}

\vspace{40pt}

{\bf Abstract}
\end{center}

We study non-supersymmetric truncations of $\omega$-deformed ${\cal N}=8$
gauged supergravity that retain a $U(1)$ gauge field and three scalars,
of which two are neutral and one charged.
We construct dyonic domain-wall and black hole solutions with AdS$_4$
boundary conditions when only one (neutral) scalar is non-vanishing,
and examine their behavior as the magnetic field and temperature of the system are varied.
In the infrared the domain-wall solutions approach either
dyonic AdS$_2 \times \mathbb{R}^2$ or else
Lifshitz-like, hyperscaling violating geometries.
The scaling exponents of the latter are $z=3/2$ and $\theta = -2$, and
are independent of the $\omega$-deformation.
New $\omega$-dependent AdS$_4$ vacua are also identified.
We find a rich structure for the magnetization of the system, including a
line of metamagnetic first-order phase transitions
when the magnetic field lies in a particular range.
Such transitions arise generically in the $\omega$-deformed theories.
Finally, we study the onset of a superfluid phase by allowing a fluctuation of the charged scalar field to condense,
spontaneously breaking the abelian gauge symmetry.
The mechanism by which the superconducting instability ceases to exist for strong magnetic fields is
different depending on whether the field is positive or negative.
Finally, such instabilities are expected to compete
with spatially modulated phases.

\thispagestyle{empty}

\pagebreak

\tableofcontents
\addtocontents{toc}{\protect\setcounter{tocdepth}{2}}


\newpage

\section{Introduction}

Holographic techniques have recently been applied to probe novel phases of matter
and materials whose unconventional behavior is tied to strong coupling, and is therefore poorly understood.
In turn these efforts have led to the discovery of new classes of
gravitational solutions and instabilities,
hinting at a rich structure of infrared (IR) phases.
We have seen the emergence of vacua characterized by a number of broken symmetries,
and scaling geometries that incorporate a dynamical critical exponent and hyperscaling violation,
an anomalous scaling of the free energy.
There is a program under way to classify such solutions and to understand
how they might arise via renormalization group (RG) flow from an ultraviolet (UV) CFT.
Bottom-up gravitational models have been used to generate a variety of these systems.
Top-down models have then served as a concrete framework for testing
these ideas, confirming many of the features
observed in the bottom-up constructions, and giving insight into the
competition between different phases.

In this paper we shall work with a non-supersymmetric but consistent
truncation of the newly discovered
one-parameter family of inequivalent ${\cal N}=8$ gauged $SO(8)$
supergravities. The ${\cal N}=8$ supergravities are characterized by
a real parameter $\omega$ that lies in the range $0\leq \omega \leq \pi/8$
\cite{Dall'Agata:2012bb,deWit:2013ija}.
The non-supersymmetric truncation we shall consider was previously
studied in
\cite{fispilwar} as a certain $SO(4)$-invariant truncation of the standard
\cite{dewitnic} undeformed ${\cal N}=8$ theory (it was called the $SO(4)'$
truncation in \cite{fispilwar}).  The truncation
involves setting to zero
certain scalar fields that play a role in restricting the range of
inequivalent values of the $\omega$ parameter in the deformed ${\cal N}=8$ theory,
and in fact after the truncation it turns out that the range of $\omega$ for
inequivalent theories is extended to $0\le\omega\le \pi/4$.
When $\omega=0$ the truncated theory can, of course, be obtained
as a truncation of the dimensional reduction of
eleven-dimensional supergravity on a seven-dimensional sphere, since
the latter gives rise to the standard $\omega=0$ supergravity.
On the other hand, the $\omega=\pi/4$ truncated theory can be obtained
via a reduction from eleven-dimensions on a
seven-dimensional Sasaki-Einstein manifold \cite{Gauntlett:2009bh}.
The higher-dimensional origin of the general $\omega$-deformed theories,
and in particular, whether they can be embedded in
eleven dimensions, is still not understood
(see \cite{deWit:2013ija} for a discussion).

The truncation we consider here retains a $U(1)$ gauge field and three scalars, of which two are neutral and one charged.
The $\omega$-deformation parameter controls the couplings of the scalars
to the gauge field, and also the structure of the scalar potential.
We shall choose the form of the gauge field so that the $(2+1)$ dimensional
UV CFT is at finite density, with chemical potential
$\mu$ and magnetic field $B$,
and then examine the possible ground states of the theory.
Throughout the paper our main interest will be in a further truncation of
the system in which only one of the neutral scalars is retained.
When the charged scalar field is also set to zero it is straightforward
to construct numerically solutions both at zero and at non-zero
temperatures,
and to examine their properties as one varies the magnetic field
and temperature in the system.

As we shall see, the behavior of the $\omega$-deformed black holes is
highly sensitive to the strength of the magnetic field.
Depending on $|B|$, as $T \rightarrow 0$ the IR geometry will approach
either a dyonic AdS$_2 \times \mathbb{R}^2$
or a solution exhibiting hyperscaling violation and Lifshitz-like scaling.
For the latter the dynamical critical exponent and the hyperscaling violating exponent are, respectively,
$z=3/2$ and $\theta=-2$, independent of the value of the $\omega$ parameter because of the
particular structure of the gauge coupling and scalar potential\footnote{There could also be purely hyperscaling violating solutions ($z=1$)
for which the electric and magnetic fields are perturbatively small, as in the neutral solutions of \cite{Charmousis:2010zz} 
and \cite{Gouteraux:2011ce} (also see \cite{Bueno:2012vx} for an investigation of scaling solutions in gauged supergravity). 
Moreover, our theory could allow for scaling solutions supported by the charged scalar field
(or a combination of the neutral and charged scalars), as in the cases studied in \cite{Gouteraux:2012yr}. For such solutions 
the scaling exponents $z$ and $\theta$ could depend on the $\omega$-deformation parameter.}.
When the magnetic field lies within a certain range, the
thermodynamically preferred black hole solutions will jump
between branches that have different IR behaviors as the temperature is changed.
This transition will then give rise to a
sudden jump in the magnetization of the system, resulting in (a line of)
first order metamagnetic phase transitions.
The latter will eventually end at a critical point -- where the phase transition
becomes second-order or higher -- when the magnetic field is tuned to a particular critical value.
Metamagnetic transitions of this type occur in a variety of materials,
including rare earth and transition metals and
strongly correlated electron systems.
Holographic studies of metamagnetism have appeared in
\cite{Lifschytz:2009sz,D'Hoker:2010rz,Bergman:2012na,Donos:2012yu}.
Here we will follow closely the analysis of \cite{Donos:2012yu}, which corresponds to
$\omega=\pi/4$ in our construction.
Finally, when the black holes are cooled down to zero temperature, they become dyonic
domain-wall solutions  interpolating between AdS$_4$ in the UV
and the two types of IR geometries that we just described.
In particular, when the deep IR is AdS$_2 \times \mathbb{R}^2$,
the system exhibits either paramagnetism or diamagnetism depending on the value of $B$.
Interestingly, for certain choices of $\omega$ the magnetic field can
be tuned so that the system will go from being paramagnetic to
diamagnetic, within the same thermodynamically preferred branch.

The well-known extensive zero temperature entropy associated with AdS$_2
\times \mathbb{R}^2$ suggests that the latter
should not describe -- generically -- the true ground state of the system.
Rather, one expects the corresponding black holes to suffer from instabilities at sufficiently low temperatures,
leading to the formation of new phases. In the extremal geometry, such instabilities would be signaled by the existence of tachyonic modes
violating the Breitenl\"ohner-Freedman (BF) bound for AdS$_2$.
Indeed, in the theories we are considering the charged scalar field can
condense, spontaneously breaking the $U(1)$ symmetry and triggering
a superfluid instability as in \cite{Gubser:2008px,Hartnoll:2008vx,Hartnoll:2008kx}.
Towards the end of the paper we shall study
the onset of this instability
in a limit in which the scalar does not back react on the
geometry. In particular, we shall determine the critical temperature at
which the instability sets in, and how it depends on the value of
the magnetic field. When $B<0$, we shall see that the charged scalar
stops condensing when the field reaches a sufficiently large value,
consistent with the Meissner effect. On the other hand, for $B>0$
the instability will stop only when the thermodynamically preferred black
hole background is hyperscaling-violating in the IR.
The asymmetry between positive and negative values of $B$ is generic
for the $\omega$-deformations of the $\omega=0$ theory.
In an appropriate range of values for $B$, we also expect to find spatially
modulated instabilities -- hinting at the presence of striped phases --
which are well known to be associated with geometries with an IR
AdS$_2 \times \mathbb{R}^2$ description \cite{Nakamura:2009tf,Ooguri:2010kt,Donos:2011bh,Donos:2011qt}.
Indeed, striped phases have been experimentally observed to compete with superconductivity
in \emph{e.g.} certain high $T_c$ superconductors \cite{SCvsStripes}.
We shall discuss briefly the range in which we expect spatially modulated instabilities to be present
and possibly dominate over the superfluid phase,
and leave a more detailed analysis to future work.

Finally, we should mention that the truncations of the $\omega$-deformed
theories that we are considering in this paper also admit new
AdS$_4$ vacua.  These have linearized instabilities within the
the full $SO(4)'$ truncation, resulting from the occurrence of
scalar flucuations whose masses lie below the Breitenl\"ohner-Freedman bound.
In one of the new AdS$_4$ vacua there is a single such unstable mode,
and the associated scalar could in fact itself be consistently truncated,
leaving a stable AdS$_4$ solution of the further-truncated theory.
It would be interesting to construct domain-wall geometries which
interpolate between two AdS$_4$ fixed points,
along the lines of \cite{Gubser:2009cg,Horowitz:2009ij}. Moreover, given the
structure of the
scalar potential and gauge couplings in our truncation -- and in particular, the fact that they generically give rise to
hyperscaling violating solutions -- we wonder whether there may be some
overlap with the construction of \cite{Bhattacharya:2014dea},
where the intermediate geometry was associated with a scaling regime.
We leave these questions to future work.

The outline of the paper is as follows. Section 2 introduces the
truncation we shall work with, and the relevant
equations of motion, while Section 3 describes the geometries that arise
in the IR.
In Section 4 we discuss the thermodynamics and construct numerically
the dyonic black hole solutions of the theory. We focus on the
behavior of the free energy and magnetization as a function of temperature.
In Section 5 we cool down our dyonic black hole solutions and
describe properties of the resulting domain-wall solutions.
We analyze superfluid instabilities triggered by the condensation of
the charged
scalar field in Section 6, and discuss the competition with striped phases.
Concluding remarks are relegated to Section 7.
Finally, Appendix A contains a brief discussion of the duality rotation
one can perform in the theory, while Appendix B
contains a description of new AdS$_4$ vacua together with a
linearized stability analysis.

\section{Non-supersymmetric $\omega$-deformed truncation}
\label{TruncationSection}

In this paper we shall study a theory obtained from the $\omega$-deformed
${\cal N}=8$ gauged supergravity by first performing a consistent
truncation to an $SO(4)$-invariant subsector of the $SO(8)$ gauged
supergravity.  This truncation, referred to as the $SO(4)'$-invariant
theory in \cite{fispilwar}, can be described conveniently in the symmetric
gauge, where the $E_7/SU(8)$ scalar coset representative of the
$SO(8)$ gauged theory is parameterised as
\be
{\cal V} = \exp\begin{pmatrix} 0 & -\fft1{2\sqrt2} \phi_{ijk\ell}\cr
             -\fft1{2\sqrt2} \phi^{mnpq} & 0\end{pmatrix}\,,
\ee
where $\phi^{ijk\ell}$ are complex scalar fields, totally antisymmetric
in the rigid $SU(8)$ indices, and obeying the complex self-duality
constraint
\be
\phi_{ijk\ell} = \fft1{4!}\, \varepsilon_{ijk\ell mnpq} \, \phi^{mnpq}\,.
\ee
Note that in the symmetric gauge $SU(8)$ and $SO(8)$ indices are identified.
Introducing coordinates $x^i$ on $\R^8$, the 35 complex
scalar fields can be written
as
\be
\Phi=\fft1{4!}\, \phi_{ijk\ell} dx^i\wedge dx^j\wedge dx^k\wedge dx^\ell\,.
\ee

  The truncation to the $SO(4)'$-invariant subsector is described in
detail in section 6 of reference \cite{fispilwar}.  For our purposes it is
convenient to view the $\R^8$ introduced above as $\C^4$, with
complex coordinates defined by
\be
z^1= x^1 + \im x^3\,,\quad z^2= x^2 + \im x^4\,,\quad
z^3 = x^5 + \im x^7\,,\quad z^4= x^6 + \im x^8\,.
\ee
The $SO(4)'$-invariant truncation in \cite{fispilwar}, which retains six
real scalar fields that we shall parameterise as
$(\lambda_0,x,\rho,\chi,\lambda_4,
  \lambda_5)$, is then given by
\bea
\Phi&=& -\fft{\lambda_0}{8\sqrt3}\,
 dz^\alpha\wedge dz^\beta\wedge d\bar z^\alpha\wedge d\bar z^\beta +
  \fft{x}{16}\, \varepsilon_{\alpha\beta\gamma\delta}\, dz^\alpha\wedge
dz^\beta\wedge d\bar z^\gamma\wedge d\bar z^\delta\nn\\
&& + \fft{\rho}{4}\,\Big[e^{\im\chi}\, dz^1\wedge dz^2\wedge dz^3\wedge dz^4
  + \hbox{c.c}\Big]\nn\\
&& +\fft{1}{48}\Big[ (\lambda_4+\im \lambda_5)\,
\varepsilon_{\alpha\beta\gamma\delta}\, dz^\alpha\wedge dz^\beta\wedge
dz^\gamma\wedge d\bar z^\delta + \hbox{c.c}\Big]\,.\label{scalarphi}
\eea
Note that the $SO(4)'$ symmetry is contained within the $SU(4)$ that
acts on $\C^4$, and so it preserves not only the $SU(4)$ invariants
$\delta_{\alpha\bar \beta}$, $\varepsilon_{\alpha\beta\gamma\delta}$
and $\varepsilon_{\bar\alpha\bar\beta\bar\gamma\bar\delta}$ of complex
geometry but also $\delta_{\alpha\beta}$ and $\delta_{\bar\alpha\bar\beta}$.
Thus we do not need to distinguished between barred and unbarred indices
in the
expression (\ref{scalarphi}) for $\Phi$, which is $SO(4)'$-invariant
but not $SU(4)$-invariant.

There is just a single $U(1)$ gauge symmetry that commutes with
$SO(4)'$, namely the $U(1)$ factor in the $U(4)=SU(4)\times U(1)$ that
acts on $\C^4$.  The surviving gauge field $A_\mu$ is embedded within
the original 28 gauge fields $A_\mu^{IJ}$ of $SO(8)$ as
\be
\ft12 A_\mu^{IJ} dx^I\wedge dx^J = \fft{\im}{2}\, A_\mu\,
   dz^\alpha\wedge d\bar z^\alpha\,.\label{gaugefield}
\ee
We can equivalently express the embedding of the gauge file as
\be
A^{IJ}_{\mu}=A_{\mu}(\sigma_0\otimes{\rm i}\sigma_2\otimes \sigma_0)^{IJ}\,.
\ee

It is evident from (\ref{gaugefield}) that the charges of the remaining
fields under the residual $U(1)$ gauge symmetry are proportional to
$(n-\bar n)$, where $n$ and $\bar n$ count the number of holomorphic
and anti-holomorphic coordinate differentials $dz^\alpha$ and
$d\bar z^\alpha$ in their expansions as differential
forms in $\C^4$.  Normalising
the charges to be $Q=\ft12(n-\bar n)$, we see from (\ref{scalarphi}) that
the scalars $\lambda_0$ and $x$ are uncharged; $\rho e^{\im \chi}$
describes a complex scalar with charge 2 and $(\lambda_4 +\im\lambda_5)$
describes a complex scalar of charge 1.

With these charge assignments we see that we can make a further
consistent truncation in which we set
\be
\lambda_4=\lambda_5=0\,,
\ee
since retained fields with charges 0 and $\pm 2$ can never act as
sources for fields of charge $\pm1$.
In terms of the notation in section 6 of \cite{fispilwar}, where the
scalar fields in the $SO(4)'$-invariant truncation were denoted
by $(\lambda_0,\lambda_1,\lambda_2,\lambda_3,\lambda_4,\lambda_5)$, the
four scalars $(\sigma,x,\rho,\chi)$ that we are retaining correspond to
\be
\lambda_0=\sigma\,,\quad
\lambda_1= \ft{\sqrt3}2\, (x- \rho \cos\chi)\,,\quad
\lambda_2=\rho \sin\chi\,, \quad
  \lambda_3= \ft12(3 x+\rho \cos\chi)\,.
\ee

   Having obtained the form of the scalar 56-bein ${\cal V}$ for the
consistent truncation we are considering, it is a mechanical, if somewhat
involved, procedure to substitute it into the expressions given in
\cite{deWit:2013ija} for the various terms in the Lagrangian of the
$\omega$-deformed ${\cal N}=8$ gauged supergravity.
We find that the $\omega$-deformed scalar potential is given by
\be
V = - f(R,x,\rho)\, \cos^2\omega - f(R^{-1},x,\rho)\, \sin^2\omega\,,
\label{pot1}
\ee
where
\be
f= \ft34 g^2 R^{-1}\, (\cosh 2 x + 3) + 3g^2 R\, \cosh x\, \cosh\rho -
    \ft12 g^2 R^3\, \sinh^2\rho\,,\label{fres}
\ee
and we have also defined
\be
R= e^{-\sigma/\sqrt3}\,.
\ee

The scalar kinetic terms are constructed as $-\ft1{48}
{\cal A}_{\mu}^{ijk\ell} {\cal A}^\mu_{ijk\ell}$, where
 ${\cal A}_\mu^{ijk\ell}$ is given by \cite{deWit:2013ija}
\be
D_\mu{\cal V}\, {\cal V}^{-1}= -\fft1{2\sqrt2} \,
\begin{pmatrix} 0 & {\cal A}_\mu^{ijk\ell}\cr
  {\cal A}_{\mu\, ijk\ell} & 0\end{pmatrix}\,.\label{cAdef}
\ee
The $SO(8)$ gauged covariant derivative of the scalar coset is defined as
\be
{\cal D}_{\mu}u_{ij}^{~~IJ}=\partial_{\mu}u_{ij}^{~~IJ}-\ft12{\cal B}_{\mu~i}^{k}u_{kj}^{~~IJ}
-\ft12{\cal B}_{\mu~j}^{k}u_{ik}^{~~IJ}-
  g(A_{\mu}^{KI}u_{ij}^{~~JK}-A_{\mu}^{KJ}u_{ij}^{~~IK})\,,
\ee
where $\ft12{\cal B}_{~j}^i$ is the composite SU(8) connection and is
determined by requiring that (\ref{cAdef}) hold.
Plugging in the ansatz, we obtain
\be
e^{-1}{\cal L}_{\rm kin} = -\ft12(\del\rho)^2 -\ft12 \sinh^2\rho\,
  (\del\chi -2gA)^2
  -\ft12 (\del\sigma)^2 -\ft32 (\del x)^2\,.
\ee

The kinetic term of the $U(1)$ gauge field is given by
\bea
e^{-1}{\cal L}_{F}&=&-\Big(\frac{e^{\sqrt3\sigma}\cos\omega-{\rm i}
\sin\omega}{\cos\omega-{\rm i}e^{\sqrt3\sigma}\sin\omega }
F^{+\mu\nu}F^+_{\mu\nu}+{\rm h.c.}\Big) \nn \\
&&\nn\\
&=& -U(\sigma)\, F^{\mu\nu} F_{\mu\nu} - W(\sigma)\,
F^{\mu\nu}\, {^*\! F}_{\mu\nu}\,,\label{gaugekinetic0}
\eea
with the gauge kinetic couplings taking the form
\beq
\label{gaugekinetic}
U(\sigma) = \frac1{\cosh\sqrt3\sigma -\cos2\omega\sinh\sqrt3\sigma}
\,,\qquad
W(\sigma) = \frac{\sin2\omega\sinh\sqrt3\sigma}{\cosh\sqrt3\sigma-
\cos2\omega\sinh\sqrt3\sigma}\,.
\eq
Combining the ingredients above, the bosonic Lagrangian for our system
becomes
\bea
\label{GeneralLagrangian}
\mathcal{L} = &-& \ft12(\del\rho)^2 -\ft12 \sinh^2\rho\,
(\del\chi -2gA)^2 -\ft12 (\del\sigma)^2 -\ft32 (\del x)^2  \nn \\
&-& U(\sigma)\, F^{\mu\nu} F_{\mu\nu} -
W(\sigma)\, F^{\mu\nu}\, {^*\! F}_{\mu\nu} - V(\sigma,x,\rho) \,,
\eea
with the scalar potential given by
\bea
\label{GeneralPotential}
V &= & - 3 g^2 [ \ft14 e^{\, \sigma/\sqrt3} \,
(\cosh 2 x + 3) + e^{-\sigma/\sqrt3} \, \cosh x\, \cosh\rho -
    \ft16 e^{-3\sigma/\sqrt3}\, \sinh^2\rho ] \cos^2\omega \nn \\
    &-&  3 g^2 [ \ft14 e^{-\sigma/\sqrt3} \, (\cosh 2 x + 3) +
e^{\, \sigma/\sqrt3}\, \cosh x\, \cosh\rho -
    \ft16 e^{3\sigma/\sqrt3} \, \sinh^2\rho ] \sin^2\omega \, .
\eea
Notice that when $x=\rho=0$ the entire $\omega$-dependence drops out of
the scalar potential.  There is still, however, $\omega$-dependence in the
coupling of the scalar field $\sigma$ to the gauge field kinetic terms.

  It is important to establish the range of the deformation parameter $\omega$
that characterises inequivalent theories.  In the full ${\cal N}=8$ gauged
supergravity, one can see that each value of $\omega$ in the line
interval $0\le \omega\le \pi/8$ describes an inequivalent theory
\cite{Dall'Agata:2012bb,deWit:2013ija}.  As is discussed there in detail, there is a symmetry under $\omega\rightarrow \omega
+\pi/2$, under which certain scalar fields undergo sign reversal
transformations.  There is also a symmetry under $\omega\rightarrow -\omega$,
combined with a parity reversal.  These two symmetries alone would imply
that inequivalent theories would correspond to points in the line interval
$0\le\omega\le \pi/4$.  However there is also another rather more
subtle symmetry in the ${\cal N}=8$ theory, under the translation
$\omega\rightarrow \omega + \pi/4$ \cite{deWit:2013ija}.  This symmetry
requires making phase transformations of some of the (complex) scalar
fields.  It is this symmetry, combined with $\omega\rightarrow -\omega$, that
results in the $0\le\omega\le \pi/8$ interval for inequivalent ${\cal N}=8$
theories.  In the truncation that we are making, the imaginary parts of
some of the complex scalars of the original ${\cal N}=8$ theory are set
to zero.  In particular, we have the real scalar field $\sigma$ that is
retained in the truncation.  As a consequence, it is no longer
possible to implement the required complex phase transformations on the
retained fields that would compensate the translation $\omega\rightarrow
\omega + \pi/4$.  The upshot is that the interval of $\omega$ corresponding
to inequivalent theories in the truncations we are considering here is
\be
0\le\omega\le \fft{\pi}{4}\,.\label{omegarange}
\ee
Note that one can indeed see from the gauge-field kinetic terms given by
(\ref{gaugekinetic0}) and (\ref{gaugekinetic}) that the theory
with $\omega=\pi/4$
is inequivalent to the theory with $\omega=0$.

Finally, the equations of motion following from (\ref{GeneralLagrangian}) are
\bea
\square \, \rho &=& \sinh\rho\, \cosh\rho\, (\del\chi-2 g A)^2
   +\fft{\del V}{\del\rho}\,,\nn\\
\square \, \sigma &=& \fft{\del V}{\del\sigma}+ \fft{\del U}{\del\sigma}\,
   F^{\mu\nu} F_{\mu\nu} +
   \fft{\del W}{\del\sigma}\, F^{\mu\nu}\, {^*\! F}_{\mu\nu}\,,\qquad
  \square \, x=\frac{1}{3}\fft{\del V}{\del x}\,,\nn\\
0&=& \nabla^\mu\Big(\sinh^2\rho\, (\del_\mu\chi - 2 g A_\mu)\Big) \,,\nn\\
0&=& \nabla^\mu\Big( U(\sigma)\, F_{\mu\nu} +
   W(\sigma)\, {^*\!\, F}_{\mu\nu}\Big) + \frac{g}{2} \sinh^2\rho\, (\del_\nu\chi- 2g A_\nu)
\,,\nn\\
R_{\mu\nu} &=& \ft12\del_\mu\rho\del_\nu\rho +
  \ft12 \del_\mu\sigma\del_\nu\sigma +\ft32 \del_\mu x \del_\nu x +
  \ft12\sinh^2\rho\, (\del_\mu\chi -2 g A_\mu)(\del_\nu\chi-2 g A_\nu)
\nn\\
&&+ 2 U(\sigma)\, (F_{\mu\rho} F_\nu^{\;\;\rho}
-\ft14 F^2\, g_{\mu\nu}) +
   \ft12 V\, g_{\mu\nu}\,.\label{eoms}
\eea

\subsection{Restricting to two scalar fields}

In the $\omega$-deformed theories discussed above it is consistent to set the neutral scalar field $x$ to zero
and retain only $\sigma$ and the charged scalar field $\rho \, e^{\, i \chi}$.
Moreover, for the cases we will consider\footnote{When $\chi=0$, the third equation in (\ref{eoms}) becomes
$\nabla^\mu (\sinh\rho^2 A_\mu) = A_\mu \, \partial^\mu(\sinh\rho^2) + \sinh\rho^2 \nabla^\mu A_\mu = 0 .$
For a gauge field with $A_r = 0$ (as it will be for us) this is satisfied when $\rho = \rho(r)$ and $\nabla^\mu A_\mu = 0$.
This is no longer true if $\rho$ depends on all coordinates.
However, we will ultimately focus on linearized perturbations of $\rho$,
i.e. $\rho = \bar\rho + \delta\rho (t,r,x,y)$, with the leading order value being $\bar\rho = 0$.
To linear order in $\delta \rho$, it is still consistent to set $\chi=0$ with
the gauge choice $ \nabla^\mu A_\mu = 0$
.} the gauge choice $\nabla^\mu A_\mu = 0$ allows to set the phase $\chi$ to zero.
With $x=\chi=0$ and $ \nabla^\mu A_\mu = 0$, the equations of motion become
\bea
\square\rho &=& 4 g^2 \sinh\rho\, \cosh\rho\, A^2
   +\fft{\del V}{\del\rho}\,,\nn\\
\square \sigma &=& \fft{\del V}{\del\sigma}+ \fft{\del U}{\del\sigma}\,
   F^{\mu\nu} F_{\mu\nu} +
   \fft{\del W}{\del\sigma}\, F^{\mu\nu}\, {^*\! F}_{\mu\nu}\,,
  \nn\\
0&=& \nabla^\mu\Big( U(\sigma)\, F_{\mu\nu} +
   W(\sigma)\, {^*\!\, F}_{\mu\nu}\Big) - g^2 \sinh^2\rho\, A_\nu
\,,\nn\\
R_{\mu\nu} &=& \ft12\del_\mu\rho\del_\nu\rho +
  \ft12 \del_\mu\sigma\del_\nu\sigma  + 2g^2 \sinh^2\rho\, A_\mu A_\nu
\nn\\
&&+ 2 \, U(\sigma)\, (F_{\mu\rho} F_\nu^{\;\;\rho}
-\ft14 F^2\, g_{\mu\nu}) +
   \ft12 V\, g_{\mu\nu}\,, \label{eomsnox}
\eea
where the scalar potential depends on the $\omega$-deformation through
\bea
\label{TwoFieldPotential}
V &= & - 3 g^2 [  e^{\, \sigma/\sqrt3}  + e^{-\sigma/\sqrt3} \, \cosh\rho -
    \ft16 e^{-\sqrt 3 \, \sigma}\, \sinh^2\rho ] \cos^2\omega \nn \\
    &-&  3 g^2 [ e^{-\sigma/\sqrt3} + e^{\, \sigma/\sqrt3}\, \cosh\rho -
    \ft16 e^{\sqrt3 \, \sigma} \, \sinh^2\rho ] \sin^2\omega \, .
\eea
Again, notice that the $\omega$-dependence cancels when $\rho = 0$.

\subsubsection{Domain-wall and black hole ansatz}

We conclude this section with the particular background ansatz which will be convenient for studying domain-wall and black hole solutions
in the rest of the paper.
We take the background to be given by
\bea
\label{ansatz}
ds^2 &=& -e^{-\beta(r)} f(r) dt^2 + \frac{dr^2}{f(r)} + r^2 d\vec{x}^2 \, , \nn \\
A &=& \phi(r) \, dt + \ft12 B (x dy - y dx) \, , \qquad \sigma = \sigma(r) \, ,
\eea
for which we have
\bea
A^2 &=& \frac{B^2}{4 r^2} (x^2 + y^2) - \frac{e^{\beta} \phi^2}{f} \, , \quad
F^2 = \frac{2B^2}{r^4} -2 e^\beta \phi^{\prime \; 2}\, , \quad
F^{\mu\nu}\, {^*\! F}_{\mu\nu} = \frac{4 \, e^{\beta/2}}{r^2} B \phi^\prime \, .
\eea
The scalar equations of motion then take the form
\bea
\square \, \rho &=&
4 g^2 \sinh\rho\, \cosh\rho\, \left[ \frac{B^2}{4 r^2} (x^2 + y^2) - \frac{e^{\beta} \phi^2}{f}\right]
   +\fft{\del V}{\del\rho}\,,\nn\\
\square \, \sigma &=& \fft{\del V}{\del\sigma}+
 \left[ \frac{2B^2}{r^4} -2 e^\beta \phi^{\prime \; 2}\right] \, \fft{\del U}{\del\sigma}
  + \left[\frac{4 \, e^{\beta/2}}{r^2} B \phi^\prime\right] \fft{\del W}{\del\sigma} \,,
\eea
while the gauge field equations of motion become
\bea
\label{first}
0 &=& f U \phi^{\prime\prime} + \phi^\prime [f U^\prime + \ft12 f U \beta^\prime + \frac{2}{r} f U ]
-g^2 \sinh\rho^2 \phi^2 - \frac{f}{e^{\beta/2} r^2} B W^\prime \, , \\
\label{second}
0 &=& - \frac{B}{2} g^2 \, y \, \sinh\rho^2 \, , \\
\label{third}
0 &=&   \frac{B}{2} g^2 \, x \, \sinh\rho^2 \, ,
\eea
where $U^\prime = \frac{\partial U}{\partial \sigma} \frac{\partial \sigma}{\partial r}$ and similarly
$W^\prime = \frac{\partial W}{\partial \sigma} \frac{\partial \sigma}{\partial r}$.

It is apparent that the choice $\rho=0$ is consistent with the equations above, and in particular with (\ref{second}) and (\ref{third}) when $B\neq 0$.
Notice that these two equations can also be satisfied by working to linear order in perturbations of $\rho$,
\beq
\label{rhoperturbation}
\rho = \bar\rho + \delta\rho \, ,
\eq
assuming that the background value is $\bar\rho=0$.
The remaining equation (\ref{first}) then fixes $\phi$.
Finally, the diagonal components of Einstein's equations are given by
\bea
\label{einsteintwofields}
&& f \beta^{\prime\prime}-f^{\prime\prime}
+ 2 \left[\frac{3}{4} f^\prime \beta^\prime +  \frac{f \beta^\prime}{r}
-\ft14 \beta^{\prime \; 2} f -  \frac{f^\prime}{r}\right] = V
- 2U\left[e^\beta \phi^{\prime \; 2} + \frac{B^2}{r^4} \right]
- 4g^2 \frac{e^\beta \sinh\rho^2}{f} \, , \nn \\
&& \rho^{\prime \, 2} + \sigma^{\prime \, 2} + \frac{2 \beta^\prime}{r} + 4g^2 \frac{e^\beta \sinh\rho^2}{f^2} \phi^2 = 0
\, , \nn \\
&& \frac{1}{r^2} \left[ 2 f + 2 r f^\prime -r f \beta^\prime \right]=
- V - 2 U \left[e^\beta \phi^{\prime \; 2} + \frac{B^2}{r^4}\right]
-\frac{g^2}{2r^2} \, B^2 \sinh\rho^2 (x^2 + y^2) \, , \nn \\
&& \frac{B^2 g^2}{f r^2} \sinh\rho^2 (x^2-y^2) = 0 \, .
\eea
The off-diagonal ($xt$, $yt$, $xy$) components are all proportional to $\sinh\rho$ and therefore
vanish trivially if $\rho=0$, as well as with the linearized perturbation (\ref{rhoperturbation}) provided again $\bar\rho=0$.
The last equation in (\ref{einsteintwofields}) is satisfied under the same conditions.
As we will see explicitly in Section \ref{superfluid instability} by working with the linearized perturbation $\delta \rho$, below a certain
critical temperature the charged scalar field $\rho$ can condense.
However, notice from (\ref{second}) and (\ref{third}) that when $B\neq 0$ the homogeneous ansatz (\ref{ansatz}) is not consistent with
a fully back-reacted solution for $\rho$. We expect that in the presence of a magnetic field the full non-linear background will be
inhomogeneous -- with a striped phase being a possible ground state.

Before closing we would like to note that
the following
transformations connect theories with different values of $\omega$,
\bea
&\omega \rightarrow -\omega &\, , \quad B \rightarrow -B\label{discrete sym 1} \, ; \\
&\omega \rightarrow \omega+\pi/2 &\, ,\quad \sigma \rightarrow -\sigma\label{discrete sym 2} \, .
\eea
When $\omega=\pi/4$, because of the combination of \eqref{discrete sym 1} and \eqref{discrete sym 2}, the theory
has the additional symmetry,
\beq
\label{additionalsymmetry}
\sigma\rightarrow -\sigma\, , \quad B\rightarrow -B \, .
\eq
Moreover, when $\omega=0$ the theory is invariant under $B\rightarrow-B$.
There is another symmetry that needs to be mentioned, which is the sign change of the vector field,
\bea\label{discrete sym 3}
\phi &\rightarrow& -\phi \, , \quad B \rightarrow -B \, ,
\ea
under which $T/\mu$ changes sign.
We will return to the role of these
transformations when we discuss the behavior of the solutions
to the $\omega$-deformed theories.

\section{The infrared geometry}
\label{IRsection}

Our main interest in this paper is in constructing domain-wall and black hole geometries with AdS$_4$ asymptotics
in the class of $\omega$-deformed SUGRA theories we have just discussed.
However, before doing so we would like to ask what types of solutions can arise
in the far infrared of the geometry, with an eye on better understanding the possible ground states of the system.
The Lagrangian we have constructed above admits a class of dyonic
AdS$_2 \times \mathbb{R}^2$ solutions, as we will show in detail below.
While these are \emph{exact} solutions, they also describe the IR of the domain wall solutions
we will construct in Section \ref{DomainWalls},
as well as the zero temperature, near-horizon limit of their non-zero
temperature generalizations -- the dyonic black holes we will
construct in Section \ref{BlackHoles}.
In addition to AdS$_2 \times \mathbb{R}^2$, we will also find zero
temperature Lifshitz-like, hyperscaling violating solutions
in the IR of the geometry. These, however, are \emph{not} exact solutions,
and break down as one moves slightly towards the UV of the geometry.
We will discuss new AdS$_4$ vacua of the $\omega$-deformed theories in Appendix \ref{AppendixAdS4}.

\subsection{AdS$_2 \times \mathbb{R}^2$ solutions}

It is evident from (\ref{eoms}) and the form of the potential (\ref{GeneralPotential})
that we can perform a consistent truncation of the theory where we set the neutral and charged scalars to zero,
\be
x=0\,,\qquad \rho=0\,,\qquad \chi=0\,.
\ee
We may then seek AdS$_2 \times \mathbb{R}^2$ solutions where we take the ansatz
\bea
ds^2 &=& - \ell^2\, r^2\, dt^2 + \fft{\ell^2\, dr^2}{r^2} + dx_1^2 + dx_2^2\,,
\nn\\
A &=& - \ell^2 E\, r dt + \ft12 B\, (x_1 dx_2 - x_2 dx_1)\,,\nn\\
\sigma&=& \sigma_0 \,,
\eea
with $\sigma_0$ denoting the constant value of the scalar.

In the vielbein basis
\be
e^0 = \ell r dt\,,\qquad e^1 = \fft{\ell dr}{r}\,,\qquad
  e^2 = dx_1\,,\qquad e^3= dx_2\,,
\ee
the non-vanishing spin connection and curvature components are given by
\be
\omega_{01}= -\ell^{-1}\, e^0\,,\qquad R_{0101}= \ell^{-2}\,,\qquad
R_{00}= - R_{11}= \ell^{-2}\,,
\ee
and the non-vanishing vielbein components of $F=dA$ are given by
$F_{01}= E$ and $F_{23}=B$.
The equations of motions then imply that
\bea
\label{sigmageneral}
\ell^{-2} &=&- V\,,\qquad E^2+B^2 = - \fft{V}{2 U(\sigma_0)}\,,\nn\\
0&=& 2 (B^2-E^2)\, U'(\sigma_0) - 4 E B \, W'(\sigma_0) + V'(\sigma_0)\,,
\eea
where a prime denotes a derivative with respect to $\sigma$.
These equations imply three conditions on the four constants of integration $E$, $B$, $\ell$ and $\sigma_0$.
It is convenient to view them as determining $E$, $B$ and $\ell$ as functions of the free parameter $\sigma_0$.

\begin{figure}[h!]
\begin{center}
\includegraphics[width=0.48\textwidth]{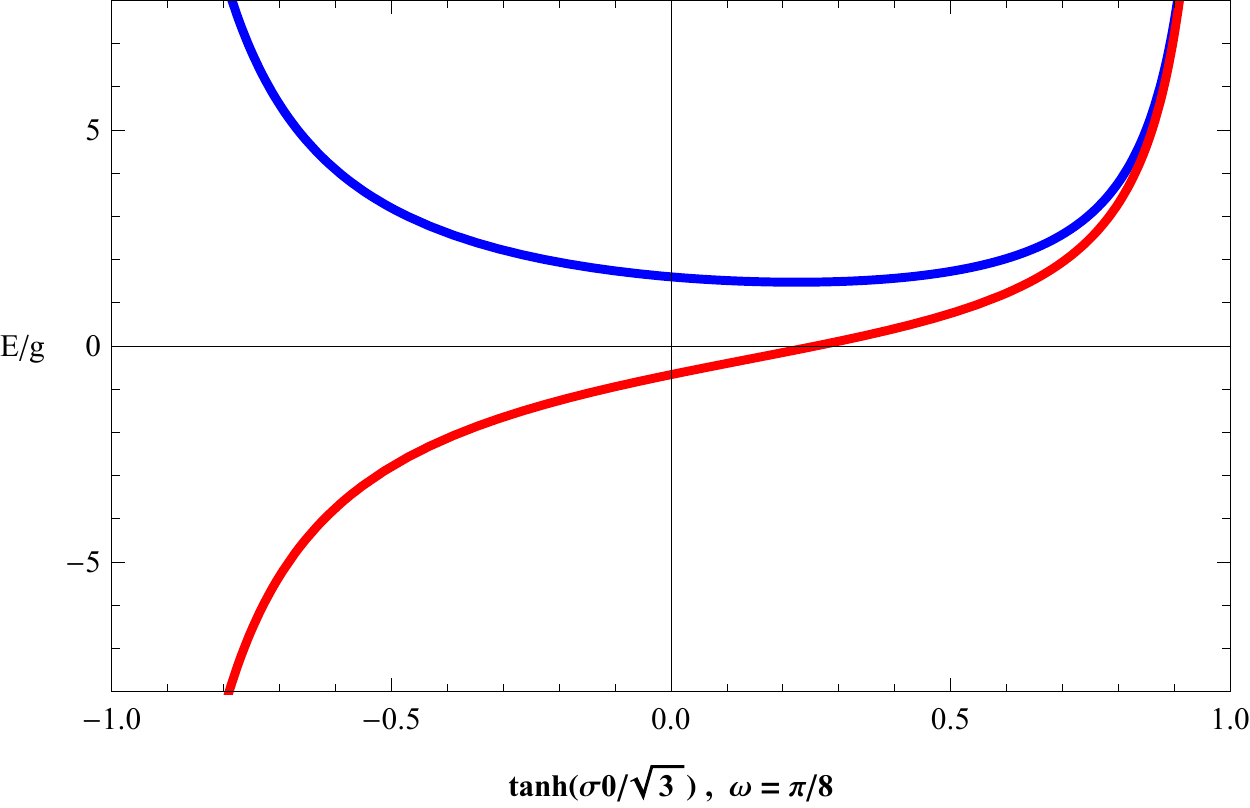}
\includegraphics[width=0.48\textwidth]{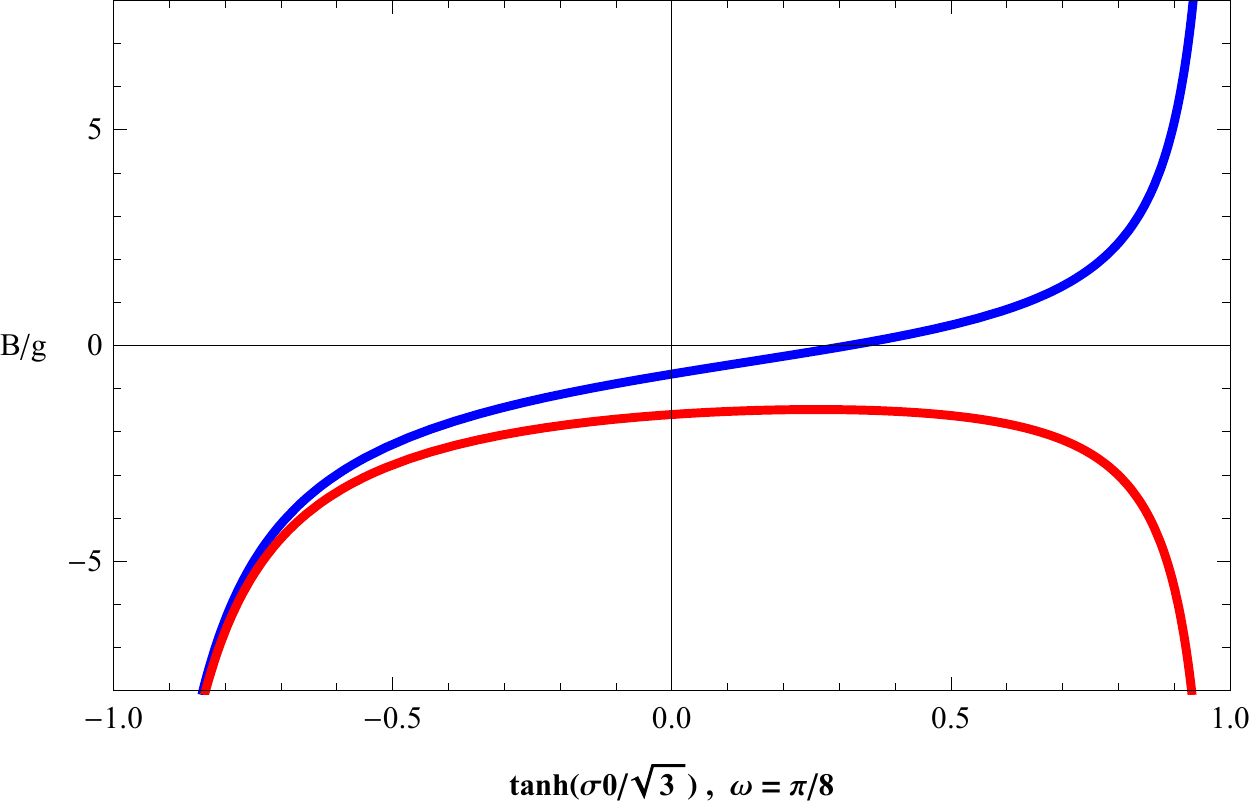}
\end{center}
\caption{\label{fig:EBfamilies}Electric (blue line) and magnetic (red line) families of solutions for $\omega = \pi/8$.
The left (right) panel shows the dependence of the electric
(magnetic) field on the
value of $\sigma_0$, expressed
as a function of $\tanh\sigma_0$ for convenience.}
\end{figure}

The dependence of $E$ and $B$ on the value of $\sigma_0$ 
for our AdS$_2 \times \mathbb{R}^2$ solutions is shown in Figure \ref{fig:EBfamilies}, where we have taken the
$\omega$-deformation parameter to be $\omega =\pi/8$.
Notice that for the blue line the electric field never vanishes (as visible from the left panel),
while for the red line it is the magnetic field which is never zero (as shown in the right panel).
For this reason, and to facilitate the comparison with \cite{Donos:2012yu},
we will refer to the class of solutions which contain the purely
electric (magnetic) AdS$_2 \times \mathbb{R}^2$ geometry
as being the \emph{electric (magnetic) family}.
In Figure \ref{fig:EBfamilies}, then, the blue line describes the electric family, while the red line
refers to the magnetic one.

As already visible from Figure \ref{fig:EBfamilies}, in our setup a dyonic
AdS$_2 \times \mathbb{R}^2$ solution is also
possible when $\sigma_0=0$, \emph{i.e.} when the scalar is not sourced.
In this case the conditions (\ref{sigmageneral}) become
\beq
\label{sigmazero}
\ell^{-2} = - V\,,\qquad E^2+B^2 = - \fft{V}{2}\,, \qquad \tan2\omega = \frac{B^2-E^2}{2EB} \, .
\eq
This is in sharp contrast with the truncation studied in \cite{Gauntlett:2009bh}, which can be obtained by setting
$\omega = \pi/4$ in our setup.
In that case $\sigma_0=0$ implied $ E B =0$,  i.e. having both electric and magnetic fields
sourced the (pseudo) scalar field. This is no longer true in the presence of a generic $\omega$-deformation --
the more complicated structure of the gauge kinetic functions (\ref{gaugekinetic}) when $\omega \neq \pi/4$
allows for a solution with vanishing $\sigma$.
As a result, the standard dyonic AdS RN black hole is a solutions to the $\omega$ deformed theories, provided the constraints
(\ref{sigmazero}) are met.

\subsection{Lifshitz and hyperscaling violating solutions}\label{hyperscaling}
\label{HVsubsection}

Next, we would like to ask whether our model supports zero temperature (non-relativistic) hyperscaling violating solutions
in the deep IR of the geometry, \emph{i.e.} at leading order in $r$ as $r\rightarrow 0$.
Using our metric parametrization (\ref{ansatz}), geometries which describe Lifshitz scaling and hyperscaling
violation take the form\footnote{The more standard parametrization is
\beq
ds^2 = - R^{\theta-2z} dt^2 + R^{\theta-2} \left( dR^2 + d\vec{x}^2 \right) \, .
\eq
}
\beq
ds^2 = - r^{\frac{2(\theta-2z)}{\theta-2}} dt^2 + r^{\frac{4}{\theta-2}} dr^2 + r^2 \, d\vec{x}^2 \, ,
\eq
and are typically supported by a running dilatonic scalar.
We consider purely magnetic solutions\footnote{The electric field diverges as one approaches the IR in these solutions.}
in which the charged scalar field is not present, and therefore take our ansatz (\ref{ansatz}) to be of the form
\bea
f(r) &=& f_0 \, r^p \, , \quad \sigma(r) = \frac{1}{\sqrt{3}}\log\sigma_0 + \kappa \log r \, , \quad
\phi(r) = 0 \, , \quad \rho(t,r,x,y)=0  \, ,
\ea
where we have anticipated that we expect the scalar to run logarithmically in the hyperscaling violation background.

An appropriate combination of Einstein's equations then takes the simple form
\beq
\sigma^{\prime \, 2} + \frac{2 \beta^\prime}{r} = \frac{\kappa^2}{r^2} + \frac{2 \beta^\prime}{r}  = 0 \, ,
\eq
whose solution is
\beq
\beta(r) = -\frac{\kappa^2}{2} \log r + C_1 \, ,
\eq
where we are allowing for an arbitrary constant.
Thus, it will be the parameters $\{\kappa, p\}$ which will determine the scaling exponents $\{z, \theta \}$
through the relations
\beq
\theta = 2- \frac{4}{p} \, , \quad z = \frac{\theta}{2} -1 + \frac{\kappa^2}{8} (2-\theta) \, .
\eq
By solving the remaining equations of motion for the system and ensuring that they are satisfied to leading order in $r$,
we find that we are forced to set (assuming for now $\kappa>0$)
\beq
\label{pkappa}
p = 1 \, , \quad \kappa = \sqrt{3}
\eq
together with
\beq
\label{Bf}
B^2 = \frac{g^2 \, \cos^2\omega}{6 \, \sigma_0^{4/3}} \, , \qquad f_0 = \frac{16 g^2}{33 \, \sigma_0^{1/3}} \, .
\eq
Note that (\ref{pkappa}) implies the following values for the scaling exponents,
\beq
\label{zthetavalues}
z= \frac{3}{2} \, , \quad \theta = -2 \, .
\eq

So far we have assumed that the scalar $\sigma$ was positive.
There is another branch of solutions on which it is negative,
\beq
\sigma(r) = -\frac{1}{\sqrt 3}\sigma_0 - \kappa \log r \, ,
\eq
with the same scaling exponents but with a rotated value for the magnetic field,
\beq
\label{negativesigma}
p=1\, , \quad \kappa=\sqrt 3\, ,  \quad B^2 = \frac{g^2 \sin^2 \omega}{6\sigma_0^{4/3}} \, ,
\quad f_0 = \frac{16 g^2}{33 \sigma_0^{1/3}} \, .
\eq

The scaling exponents (\ref{zthetavalues}) are the same as the ones found in \cite{Donos:2012yu} for the SUGRA truncation studied in
\cite{Gauntlett:2009bh}, which corresponds to taking $\omega = \frac{\pi}{4}$  and $g=2$.
The main difference here is the $\omega$ dependence appearing in (\ref{Bf}) and (\ref{negativesigma}).
Recall that in our model the couplings of the scalar $\sigma$ to the gauge field and the scalar potential
depend on various combinations of exponentials of the form $e^{\pm\sqrt{3}\sigma}$ as well as $e^{\pm\frac{\sigma}{\sqrt{3}}}$,
whose arguments do not depend on $\omega$.
Since it is precisely the structure of the \emph{argument} of the exponentials that fixes the scaling exponents,
our construction unfortunately does not allow for $\omega$-dependent values for $\{z,\theta\}$.

\section{Black hole solutions}
\label{BlackHoles}

We are now ready to study the behavior of dyonic black hole solutions in the $\omega$-deformed theories described by
(\ref{GeneralLagrangian}), for the case in which the neutral and charged complex scalars vanish, $x = \rho = \chi = 0$.
We are interested in geometries which are asymptotic to AdS$_4$, so that the dual gauge theory will be a CFT in $2+1$ dimensions.
We also want the latter to be at finite density, with chemical potential $\mu$ and magnetic field $B$.
We work with the background ansatz (\ref{ansatz}), which we include again here for convenience,
\bea
\label{ansatzagain}
ds^2 &=& -e^{-\beta(r)} f(r) dt^2 + \frac{dr^2}{f(r)} + r^2 d\vec{x}^2 \, , \nn \\
A &=& \phi(r) \, dt + \ft12 B (x dy - y dx) \, , \qquad \sigma = \sigma(r) \, ,
\eea
and we start constructing the solutions by writing down the expansions for
the geometry about the boundary and the horizon.\\
\noindent
\underline{{\bf UV expansion}}\\
At the boundary, as $r\rightarrow \infty$, the metric should approach
AdS$_4$ and the system should be at finite density.
We solve the equations of motion perturbatively in $1/r$ (setting $G=1/16\pi$)
and find the following expansion,
\bea
f&=& g^2 r^2+\frac{g^2\sigma_1^2}4-\frac1{2r}(\varepsilon-\ft43\sigma_1\sigma_2)+\ldots,\nn\\
\beta&=&\beta_0+\frac{\sigma_1^2}{4r^2}+\frac{2\sigma_1\sigma_2}{3r^3}+\ldots,\nn\\
\phi&=&e^{-\beta_0/2}\Big(\mu-\frac{q}{r}-\frac{\sqrt{3}B\sigma_1\sin2\omega-\sqrt{3}q\sigma_1\cos2\omega}{2r^2}\Big)+\ldots,\nn\\
\sigma&=&\frac{\sigma_1}{r}+\frac{\sigma_2}{r^2}+\frac{5\sigma_1^3}{72r^3}+\ldots \, .
\label{uvexp}
\eea
The parameters $\sigma_1$ and $\sigma_2$ represent, respectively, the source and VEV
of the operator $\mathcal{O}_\sigma$ dual to the scalar.
In our numerics we will turn off the source $\sigma_1$ for $\mathcal{O}_\sigma$  and take its
scaling dimension to be $\Delta = 2$, corresponding to a relevant deformation of the UV CFT.
Thus, taking into account the choice $\sigma_1 = 0$, we have six parameters describing the UV expansion of the geometry,
$\{\varepsilon, \sigma_2, \beta_0, \mu, q, B \}$.

\noindent
\underline{{\bf IR expansion}}\\
The near-horizon $r \sim r_+$ behavior of the background takes the form
\bea
f&=&f_+(r-r_+)+\ldots,\nn\\
\beta &=&\beta_{+}+\ldots,\nn\\
\phi&=&\phi_{+}(r-r_+)+\ldots,\nn\\
\sigma&=&\sigma_{+}+.\ldots  \, ,
\label{irexp}
\eea
described by the four parameters $\{ \beta_{+},\phi_{+},\sigma_{+},r_+ \}$, with
\be
f_{+}=3r_{+}g^2\cosh\frac{\sigma_{+}}{\sqrt{3}}
-\frac{B^2+\phi_{+}^2r^4_{+}e^{\beta_{+}}}{r^3_{+}(\cosh\sqrt{3}\sigma_{+}-\cos2\omega\sinh\sqrt{3}\sigma_{+})}.
\ee
The higher order terms in the IR expansion are somewhat involved and will therefore not be included here.
The numerical analysis will take them into account.

Similarly to \cite{Donos:2012yu}, the equations of motion for our dyonic black hole ansatz
are invariant under the two scaling symmetries
\bea
&&t\rightarrow \lambda t,\quad e^{\beta}\rightarrow \lambda^2 e^{\beta},\quad \phi\rightarrow \lambda^{-1}\phi \,  ;\nn\\
&&r\rightarrow r,\quad (t,x,y)\rightarrow \lambda^{-1}(t,x,y),\quad f\rightarrow \lambda^2 g,\quad \phi\rightarrow \lambda\phi,\quad B\rightarrow \lambda^2 B.
\eea
After solving the full set of equations numerically by integrating out from the horizon to infinity,
we will use the first scaling symmetry to fix $\beta_0=0$.
We will use the second scaling symmetry to identify the scale-invariant quantities
$T/\mu$, $B/\mu^2$ with which we will label inequivalent solutions.


\subsection{Thermodynamics}

Following the discussion of \cite{Gauntlett:2009bh}, we analytically continue by setting $t=-i\tau$, and $I=-iS$.
We can then obtain two expressions for the on-shell action for the class of the solutions we are studying.
The first expression is given by the integral of a total derivative
\be
I_{\rm bulk}=\frac{\Delta\tau {\rm vol}_2}{16\pi G}\int_{r_+}^{\infty} dr[r^2e^{-\beta/2}(f'-f\beta'-4U(\sigma)e^{\beta}\phi\phi')+4BW(\sigma)\phi]',
\label{euaction1}
\ee
where ${\rm vol}_2 \equiv\int dx^1dx^2$.
The second can be written as
\be
I_{\rm bulk}=\frac{\Delta\tau {\rm vol}_2}{16\pi G}\int_{r_+}^{\infty} dr\{[2rfe^{-\beta/2}]'+4B^2r^{-2}e^{-\beta/2}U(\sigma)+4BW(\sigma)\phi'\}.
\label{euaction2}
\ee
The total action includes the Gibbons-Hawking surface term supplemented by counterterms
\be
I_{\rm tot}=I_{\rm bulk}+I_{\rm surf}+I_{\rm ct},
\ee
where
\bea
I_{\rm surf} &=&
 \frac{1}{8\pi G}\int_{\partial {\cal M}}d\tau d^2x \sqrt{h}K\,\label{action},\nn\\
I_{\rm ct}&=&\frac{1}{8\pi G }\int_{\partial {\cal M}} d\tau d^2x
 \sqrt{h}\Big(\frac2{\ell}+\frac{\ell}2{\cal R}\Big)
+\frac{1}{48\pi G} \int_{\partial {\cal M}} d^3x \sqrt{h}
\Big(\sigma n^{\mu}\partial_{\mu}\sigma-\frac1{2\ell}\sigma^2
\Big)\,.
\label{counterterm}
\eea
In the equation above $K_{\mu\nu}\equiv -\ft12 (\nabla_{\mu}n_{\nu}+\nabla_{\nu}n_{\mu})$ is the extrinsic
curvature of the boundary surface, with $n_{\mu}$ being the outward
unit normal vector. The curvature radius of AdS is $\ell=1/g$, and
${\cal R}$ is the Ricci scalar of the boundary metric.
It should be mentioned that here we use the counterterms given
in \cite{Lu:2013ura}, which are
different from the ones used in \cite{Donos:2012yu}.
The reason is that the counterterms chosen by \cite{Donos:2012yu} apply
specifically to the cases where the scalar
$\sigma$ satisfies Dirichlet or Neumann boundary conditions,
corresponding to $\sigma_1=0$ or $\sigma_2=0$.
However, from the dyonic black holes found in \cite{Lu:2013ura}, one can see that the system studied here also
admits black hole solutions in which $\sigma$ satisfies the mixed boundary condition $\sigma_2\propto \sigma_1^2$
corresponding to turning on a triple trace deformation in the dual theory.
It was shown in \cite{Hertog:2004dr} that the boundary condition $\sigma_2\propto \sigma_1^2$ preserves all
the asymptotic AdS symmetries, therefore the holographic stress tensor should be traceless in this case.
As we show below, while the stress tensor calculated using the boundary
term \cite{Lu:2013ura} has such a property.\footnote{More generally,
we can parameterise with constants $\alpha$ and $\beta$ a family of
counterterms
\bea
I_{ct} =\frac{1}{8\pi G }\int_{\partial {\cal M}} d\tau d^2x
 \sqrt{h}\Big(\frac2{\ell}+\frac{\ell}2{\cal R}
+\ft12 (1-\alpha) \sigma n^\mu\del_\mu\sigma +
  \fft{1-2\alpha}{4\ell}\, \sigma^2 +
   \fft{\beta}{3\ell}\, \sigma^3 \Big)\nn
\eea
that give rise to finite expressions for the renormalised action, stress
tensor and mass.  Specifically, we find that the trace of the stress tensor
and the mass are given by
\bea
\tau^\mu{}_\mu = (2\alpha-\ft43) g^2 \sigma_1\sigma_2 +\ft43 g^2 \beta
       \sigma_1^3\,,\qquad
M= M_{AMD} + (\ft23-\alpha) g^2 \sigma_1\sigma_2 - \ft23 g^2\beta
\sigma_1^3\,.\nn
\eea
Our choice, $\alpha=\ft23,\ \beta=0$, gives a traceless stress tensor
and $M=M_{AMD}$, the AMD value for the mass. An alternative choice
would be $\alpha=1,\ \beta=0$, for which $\sigma_1$ would acquire
a more standard holographic interpretation as a source term $J$.  Any
assignment of values for $\alpha$ and $\beta$ would give a valid definition
of an ``energy'' of a black hole, with one differing from another by
a Legendre transformation.  Our preference for the present purposes of
discussing black-hole thermodynamics is to choose the energy functional
that coincides with the AMD definition, and which gives a trace-free
boundary stress tensor.  Since, by contrast, our holographic discussions
are all
concerned with solutions where $\sigma_1=0$, for which the
thermodynamic quantities are then independent of $\alpha$ and
$\beta$, the choice of counterterm
becomes immaterial for those purposes.}

  Using these counterterms, the renormalized energy-momentum tensor
is given by $T^{\mu\nu}\equiv (2/\sqrt{-h})\, \delta I/\delta h_{\mu\nu}$,
yielding
\be
T_{\mu\nu}=\frac1{8\pi G}
\Big(K_{\mu\nu}-Kh_{\mu\nu}-\frac2{\ell}h_{\mu\nu}+
\ell \, ({\cal R}_{\mu\nu} -\ft12 {\cal R}\, h_{\mu\nu})
+\frac16 h_{\mu\nu}(\sigma n^{\rho}\partial_{\rho}\sigma-\frac1{2\ell}\sigma^2)
\Big)\,.
\ee
The stress tensor $\tau^i_{~j}$ of the dual boundary theory can be calculated as
\be
\tau^i_{~j}=2r^3T^i_{~j}|_{r=\infty},
\ee
from which we find
\be
\tau^{t}_{~t}=-\varepsilon,\quad \tau^{x}_{~x}=\tau^{y}_{~y}=\frac{\varepsilon}{2}.
\ee
We notice that the energy density coincides with the AMD mass density \cite{Ashtekar:1999jx}, and the stress tensor of the dual CFT is
traceless as expected.

Next, we define the thermodynamic potential $W\equiv T[I_{\rm tot}]\equiv {\rm vol}_2 w$, where
the temperature of the black hole is given by
\be
T=\frac{e^{(\beta_0-\beta_+)/2}f_+}{4\pi}.
\ee
Using the expression (\ref{euaction1}) and the expansions (\ref{uvexp}) and (\ref{irexp}), we obtain
\be
w=\varepsilon-4\mu q-Ts \, ,
\label{freeenergy}
\ee
where the $s$ is the entropy density given by
\be
s=4\pi r^2_{+} \, .
\ee
On the other hand, making use of (\ref{euaction2}), we obtain
\be
w=-\frac{\varepsilon}2+4e^{\beta_0/2}\int_{r_+}^{\infty} dr \{B^2r^{-2}e^{-\beta/2}U(\sigma)+BW(\sigma)\phi' \}.
\ee
Equating this expression with (\ref{freeenergy}) would give a Smarr type formula.
Following the Wald procedure, the first law of thermodynamics takes the form
\be
\delta\varepsilon=T\delta s+4\mu\delta q-m\delta B+\ft13g^2(\sigma_1\delta\sigma_2-2\sigma_2\delta \sigma_1),
\ee
where $m$ is the magnetization per unit volume
\be
m=-4e^{\beta_0/2}\int_{r_+}^{\infty} dr \{Br^{-2}e^{-\beta/2}U(\sigma)+W(\sigma)\phi'\}.
\ee

Finally, using the formulae given in Appendix \ref{AppendixDuality}, we would like to show how the free energy
of the $\omega$-deformed theory differs from that of the undeformed theory.
Recall that the $U(1)$ field strength in the $\omega$-deformed theory is related to the undeformed one by a duality rotation,
\be
F_{\omega}=\cos\omega F_0-\sin\omega e^{\sqrt{3}\sigma} *F_0 \, .
\ee
Plugging in the dyonic black hole ans\"{a}tz for $F$,
\be
F=-\phi'dt\wedge dr+Bdx\wedge dy \, ,
\ee
we obtain
\be
\phi'_{\omega}=\cos\omega\phi'_0+\sin\omega e^{\sqrt{3}\sigma}e^{-\beta/2}B_0/r^2,
\quad B_{\omega}=\cos\omega B_0-r^2e^{\beta/2}e^{\sqrt{3}\sigma}\sin\omega\phi'_0 \, .
\ee
Using the UV expansion of the fields (\ref{uvexp}), we can derive
\be
q_{\omega}=\cos\omega q_0+\sin\omega B_0,\quad B_{\omega}=\cos\omega B_0-\sin\omega q_0,\quad
\mu_{\omega}=\cos\omega \mu_0-\ft14\sin\omega m_0.
\ee
Under the duality rotation the energy, temperature and entropy of the black hole solutions do not change, but
the free energy does, so that
\be
w_{\omega}=\varepsilon-4\mu_{\omega}q_{\omega}-Ts \;\; \neq \;\; w_0=\varepsilon-4\mu_0 q_0-Ts \, .
\ee

\subsection{Magnetic Field Induced Transitions}
\label{MagnFieldSubsection}

We have constructed dyonic black hole solutions to our $\omega$-deformed theories numerically by
building on the UV and IR expansions  (\ref{uvexp}) and (\ref{irexp}).
The solutions have AdS$_4$ asymptotics, to ensure that in the UV the dual theory is described by a three-dimensional CFT.
Moreover, we have taken the source $\sigma_1$ for the operator $\mathcal{O}_\sigma$ dual to $\sigma$ to be zero
-- to avoid deforming the UV CFT -- and chosen its conformal dimension to be $\Delta = 2$, corresponding to a relevant perturbation.
We expect to find a rich phase structure as one varies the magnetic field and temperature in the system.
In particular, from the analysis of \cite{Donos:2012yu} (corresponding to the deformation choice $\omega = \pi/4$)
we expect to find a line of first order metamagnetic transitions when the magnetic field becomes sufficiently large.
Indeed, this will be a generic feature of our $\omega$-deformed theories.
Moreover, as we cool the thermodynamically preferred black holes down to zero temperature,
the resulting domain-wall solutions will approach either AdS$_2 \times \mathbb{R}^2$ or a hyperscaling violating solution in the IR,
depending on the strength of $B$.
To facilitate the comparison with \cite{Donos:2012yu}, we will adopt their notation from here on.

\begin{figure}[h!]
\begin{center}
\includegraphics[width=0.6\textwidth]{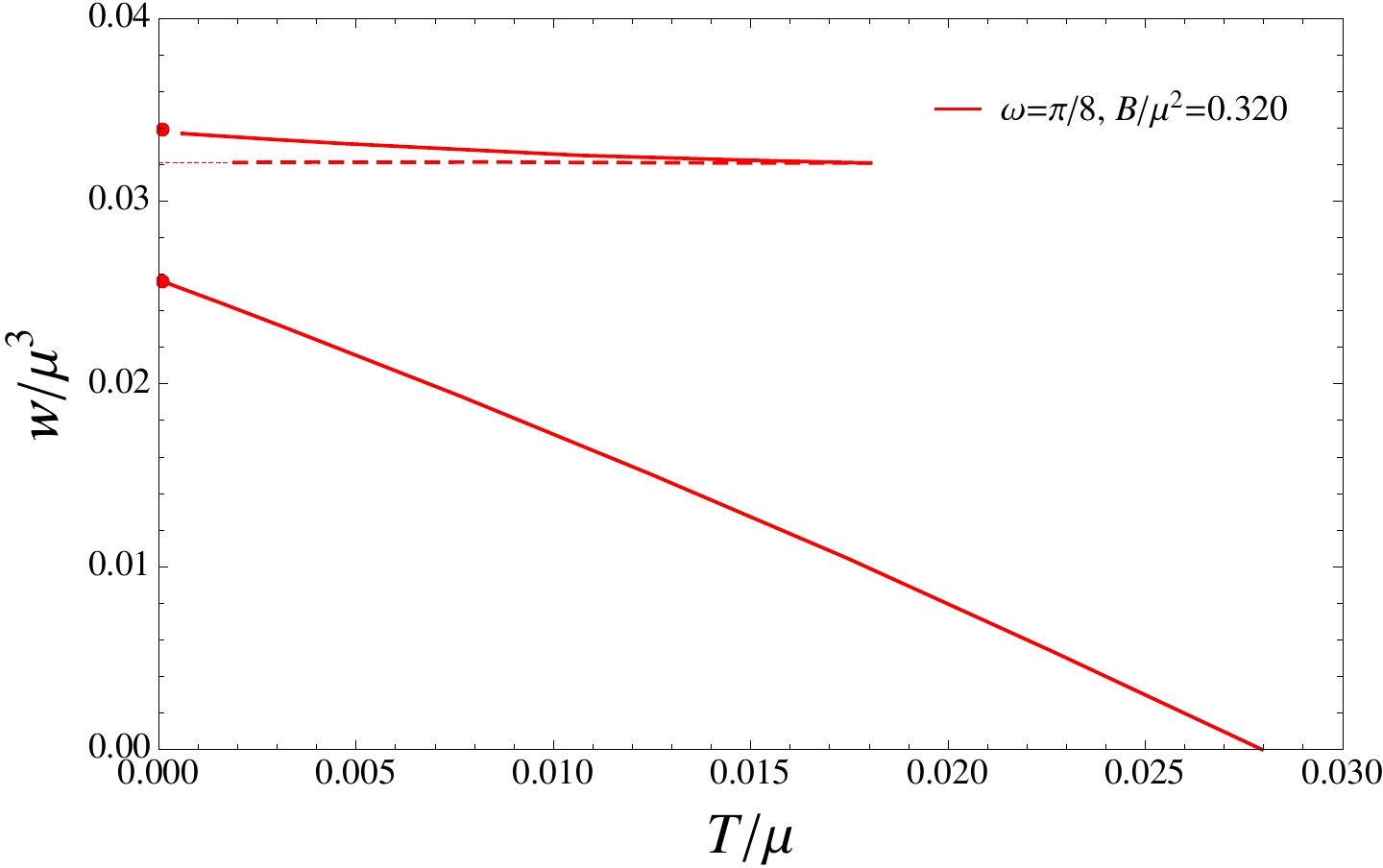}
\end{center}
\caption{Typical plot of the free energy as a function of temperature for $\omega=\pi/8$ and $g=2$ when the magnetic field
is in the range (\ref{Brange}). The solid lines describe black holes whose zero temperature limits are domain-walls
approaching in the IR AdS$_2 \times \mathbb{R}^2$ (denoted by dots). The thick dashed line describes black holes whose domain-wall limit
approach hyperscaling violating solutions in the IR. The thin dashed line is a naive extrapolation of the numerical data (thick dashed line)
to very low temperatures.
\label{fig:blackholetypical}}
\end{figure}

Let's start by discussing how the temperature dependence of the free energy is affected by the magnetic field.
In Figure \ref{fig:blackholetypical} we show a typical plot of the free energy as a function of
temperature for a moderately low value of $B$, which we take to be in the range\footnote{When $B<0$
we denote the corresponding range by
$\left(\frac{B}{\mu^2}\right)_{I}^{neg} <  \frac{B}{\mu^2} < 0$.}
\beq
\label{Brange}
0 <  \frac{B}{\mu^2}  < \left(  \frac{B}{\mu^2}  \right)_I \, ,
\eq
with $ \left( B/\mu^2\right)_I$
to be defined shortly.
We have chosen $\omega = \pi/8$, but the structure seen in the figure is insensitive to the specific value
of the $\omega$ parameter.
We find three branches of solutions, only one of which can seemingly be heated up to arbitrarily high temperatures.
The solid lines describe black hole geometries whose zero temperature limit are domain-walls approaching
dyonic AdS$_2 \times \mathbb{R}^2$ solutions in the IR. The latter belong to the \emph{electric family}\footnote{We expect the magnetic family of
solutions to have higher free energy, as we will discuss in Section \ref{DomainWalls}.} we described in Section \ref{IRsection}
and are denoted by dots in the plot.
On the other hand, the (thick) dashed line describes a black hole whose zero temperature, deep IR limit is
a hyperscaling violating solution\footnote{We have
verified numerically that the scaling of the entropy with temperature as we approach $T\sim 0$ matches that of
a Lifshitz geometry with hyperscaling violation, \emph{i.e.} $s(T)\sim T^{8/3}$ when the exponents
are $z=3/2$ and $\theta=-2$.} of the type discussed in Section \ref{HVsubsection}.
In all the figures in this section the thin dashed lines are a naive extrapolation of the numerics (thick dashed lines)
to very low temperatures.

\begin{figure}[h!]
\begin{center}
\includegraphics[width=0.49\textwidth]{figures/bhpiover82.pdf}
\includegraphics[width=0.49\textwidth]{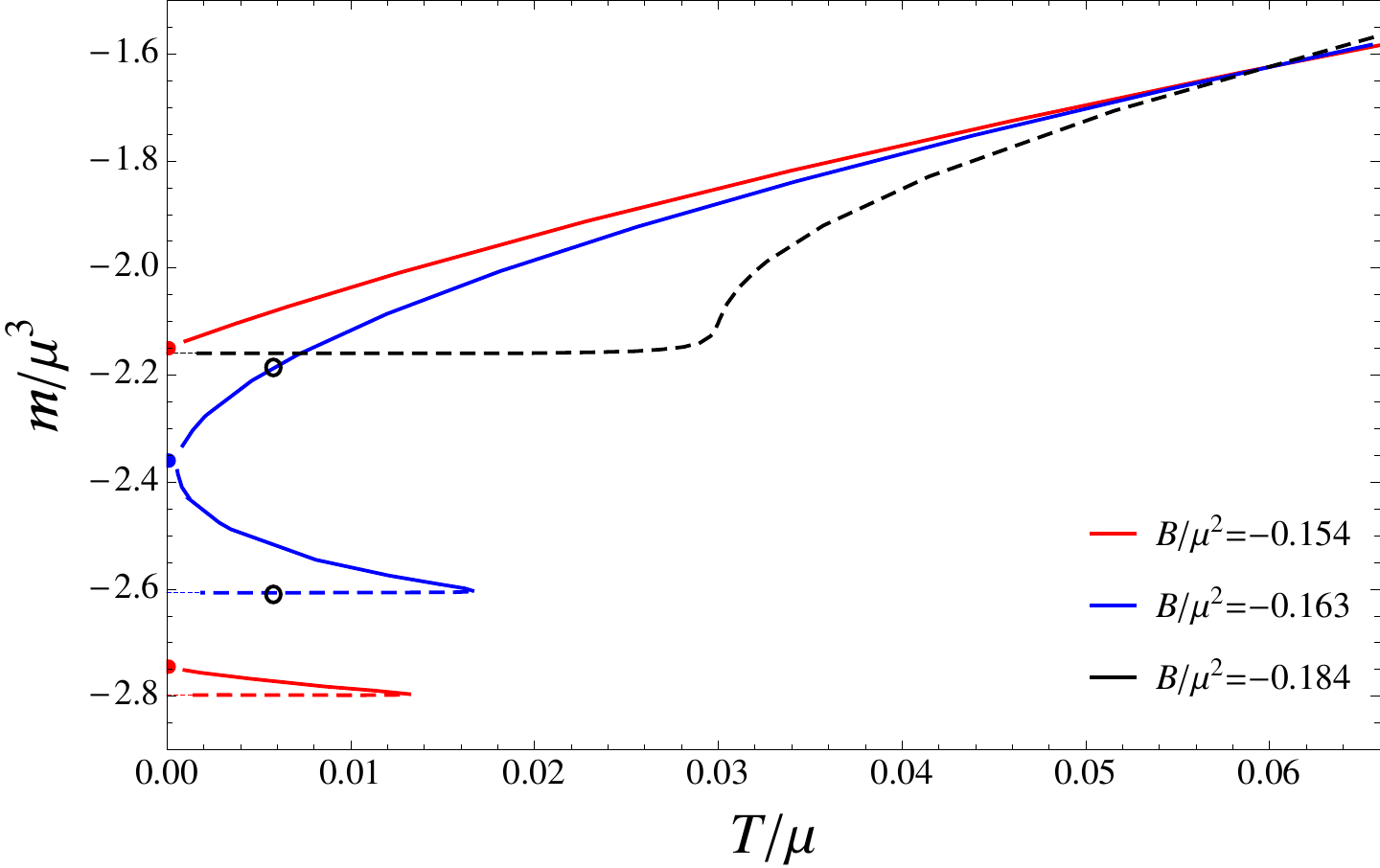}
\end{center}
\caption{Magnetization as a function of temperature for $\omega=\pi/8$ and $g=2$, for different values of $B/\mu^2$.
In the left (right) panel the magnetic field is positive (negative).
The solid/dashed lines and the dots are the same as in Figure \ref{fig:blackholetypical}.
The open circles mark the point at which the magnetization
undergoes a sudden jump, denoting a first order metamagnetic phase transition.
\label{fig:BlackHolePhaseDiagramMagnetization}}
\end{figure}

We see from the figure that there are two distinct dyonic AdS$_2 \times \mathbb{R}^2$ geometries at $T=0$ (denoted by the two dots).
The black hole branch which is thermodynamically preferred is the one whose temperature can be arbitrarily high,
and approaches the AdS$_2 \times \mathbb{R}^2$ with the lower free energy.
The phase structure shown in Figure \ref{fig:blackholetypical} turns out to be typical
as long as the magnetic field is in the range (\ref{Brange}).
In this discussion we will assume that $B>0$ but the same argument goes through for a negative field.
The value
$ \frac{B}{\mu^2}  = \left(\frac{B}{\mu^2}\right)_I$
(which is $\omega$-dependent) is defined to be such that, at zero temperature,
the bottom AdS$_2 \times \mathbb{R}^2$ solution overlaps with the hyperscaling violating one --  the two geometries have the same free energy.
Thus, if the magnetic field is any larger, at very low temperatures the hyperscaling violating branch becomes
thermodynamically preferred.
The temperature dependence of the magnetization $m$ when the field is within the range (\ref{Brange}) is shown by the red curve
in Figure \ref{fig:BlackHolePhaseDiagramMagnetization}.
In the left panel the thermodynamically preferred black hole branch corresponds to the bottom (red) curve,
along which $B>0$ and $m$ becomes more negative as the temperature is raised.
Thus, the system is becoming more ordered as it is heated, with the magnetization opposing the direction of the magnetic field.
On the other hand in the right panel it is the top (red) curve which is favored,
along which both $B$ and $m$ are negative, and the latter becomes less negative as $T$ increases.
In this case the magnetization is aligned with the field, and the system becomes less ordered as the temperature is raised.

\begin{figure}[h!]
\begin{center}
\includegraphics[width=0.49\textwidth]{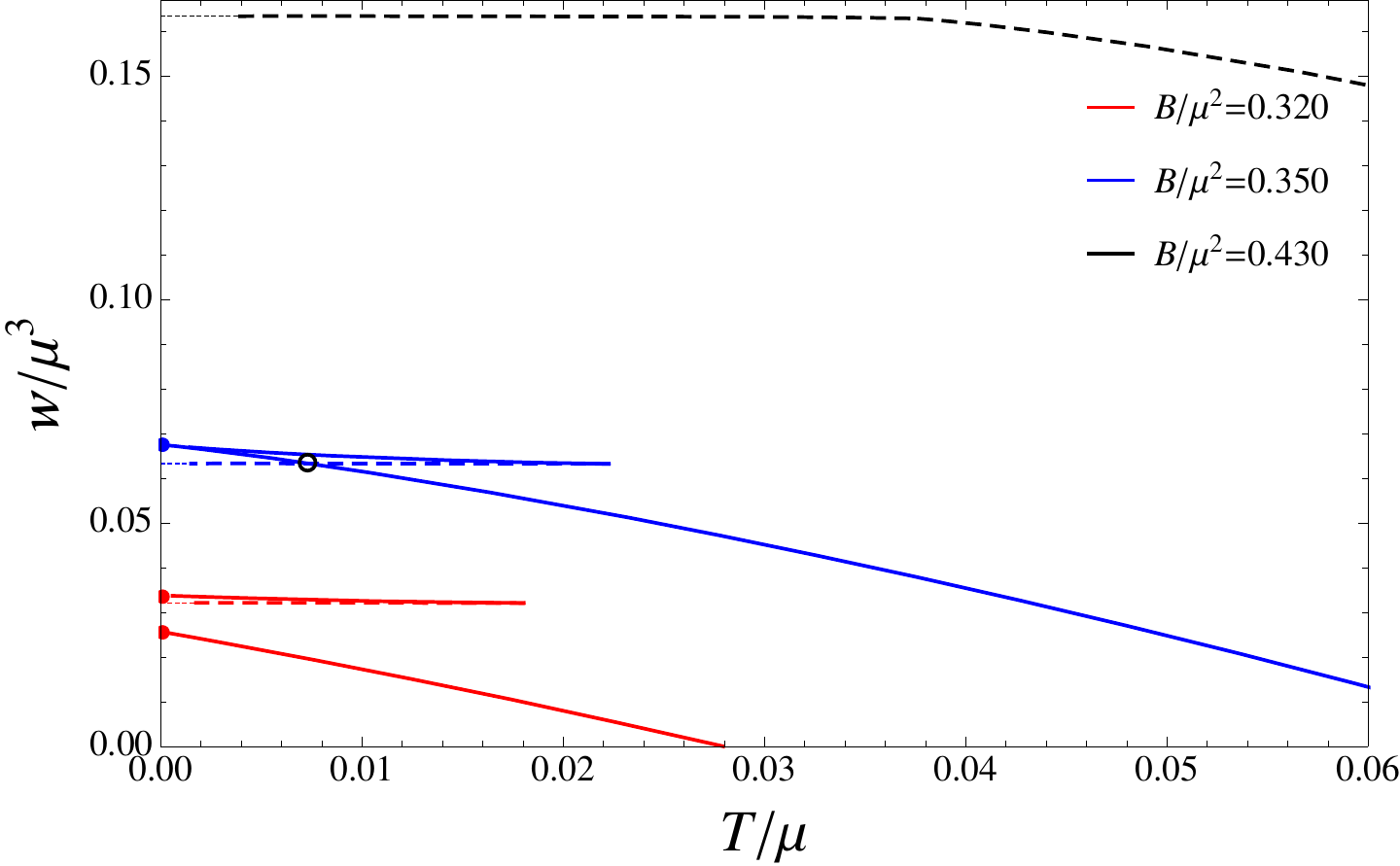}
\includegraphics[width=0.49\textwidth]{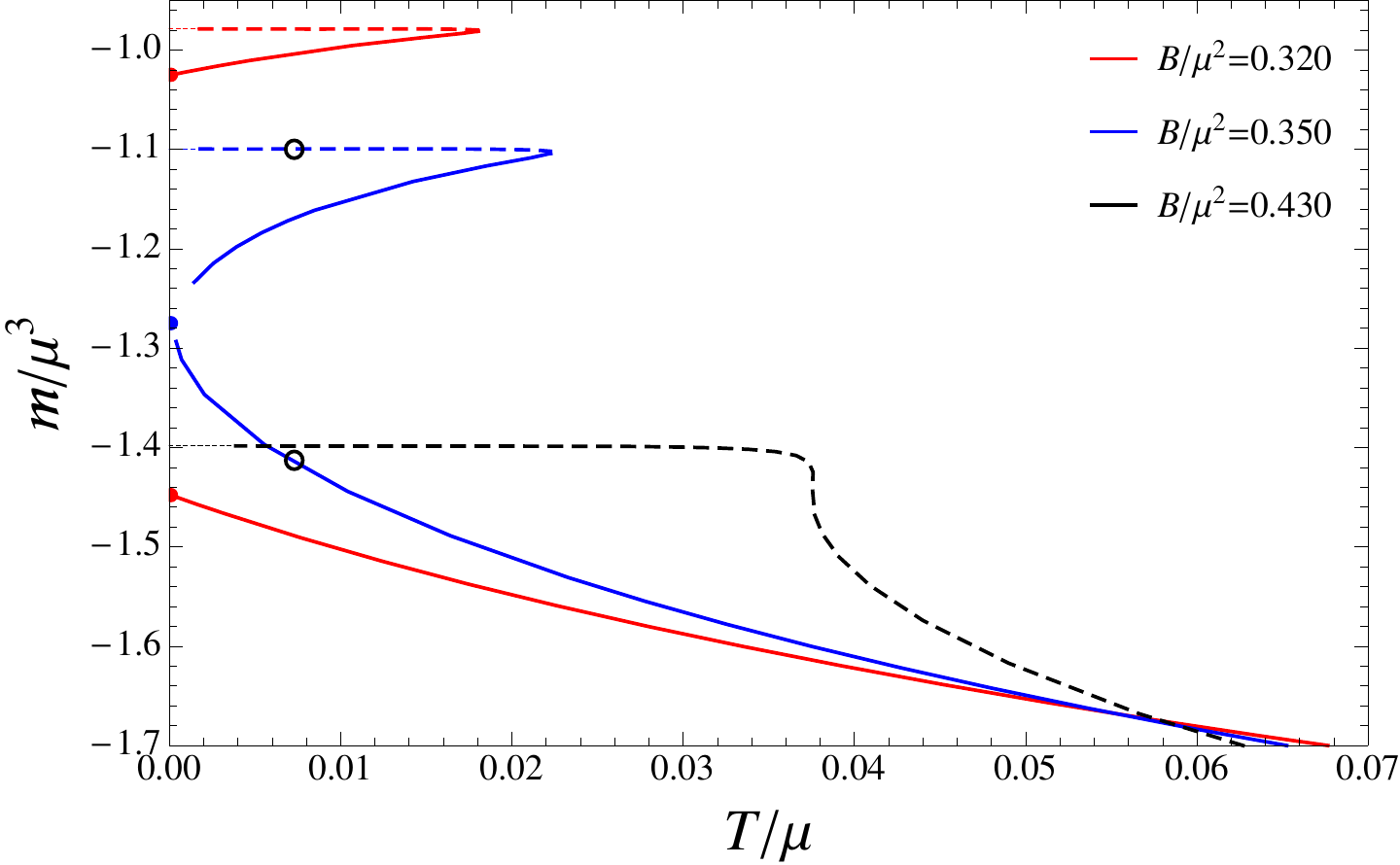}
\end{center}
\caption{Free energy as a function of temperature for $\omega=\pi/8$ and $g=2$, for different values of $B/\mu^2$.
In the left (right) panel the magnetic field is positive (negative), and $B$ increases (decreases) from bottom to top.
The solid/dashed lines and the dots are the same as in Figure \ref{fig:blackholetypical}.
The open circles mark the point at which the thermodynamically preferred branch switches from the one with
AdS$_2 \times \mathbb{R}^2$ in the IR to the hyperscaling violating branch.
\label{fig:BlackHolePhaseDiagram}}
\end{figure}

As $B/\mu^2$ is raised above $\left(B/\mu^2\right)_I$, the two zero temperature AdS$_2 \times \mathbb{R}^2$ solutions become closer to each other,
and overlap when the ratio reaches a critical value\footnote{The analogous critical value for $B<0$ will be denoted by
$\left( B/\mu^2\right)_{max}^{neg}$.} which we denote by $\left(B/\mu^2\right)_{max}$.
At this point there is only a single domain-wall solution whose IR is a dyonic
AdS$_2 \times \mathbb{R}^2$ geometry
belonging to the electric family.
This behavior is visible in Figure \ref{fig:BlackHolePhaseDiagram}, where we plot the free energy as a function of
temperature for increasing values of magnetic field and for $\omega=\pi/8$.
In the left panel we have taken $B>0$ and the field increases from bottom to top.
On the other hand in the right panel $B<0$ and becomes more negative from top to bottom.
In Figure \ref{fig:BlackHolePhaseDiagram} the overlap of the two $T=0$
AdS$_2 \times \mathbb{R}^2$ geometries
occurs for the choice of magnetic field shown in the blue curve,
i.e. $\left( B/\mu^2\right)_{max} = 0.350$
in the left panel and $\left( B/\mu^2\right)_{max}^{neg} = -0.163$
in the right panel.
Thus, when the field is such that\footnote{When the field is $B<0$ the range becomes
$\left( \frac{B}{\mu^2} \right)_{max}^{neg} \leq \frac{B}{\mu^2} < \left( \frac{B}{\mu^2} \right)_{I}^{neg}$.}
\beq
\label{Brange2}
\left( \frac{B}{\mu^2} \right)_{I} < \frac{B}{\mu^2} \leq \left( \frac{B}{\mu^2} \right)_{max} \, ,
\eq
the branch which is thermodynamically favored at low temperatures is the hyperscaling violating one.
As the temperature is raised, we eventually cross over to the black hole branch with an associated $T=0$ AdS$_2 \times \mathbb{R}^2$ IR description,
which dominates at sufficiently high temperatures.
We emphasize that this behavior is typical when the magnetic field is within the range (\ref{Brange2}).

The transition between the two branches is \emph{first order}, as visible from the cusp in the free energy plot where the two
lines meet.
This is also confirmed by the behavior of the magnetization,  displayed in Figure \ref{fig:BlackHolePhaseDiagramMagnetization}.
As we move from one black hole branch to the other by following the blue curve in the figure,
the magnetization suffers a sudden, discontinuous jump, i.e. it undergoes a \emph{metamagnetic} first order phase transition.
This was already seen in \cite{Donos:2012yu} for the particular $\omega = \pi/4$ case, and is generic
in our $\omega$-deformed theories.

Metamagnetic transitions occur in a number of materials \cite{Ito,Kaczmarsca}, including
strongly correlated electrons systems.
An example is the layered ruthenate metal $Sr_3 Ru_2 O_7$, which exhibits -- for a sufficiently large value of
magnetic field -- a line of first order, non-zero temperature metamagnetic phase transitions which end at a critical point.
In the particular $Sr_3 Ru_2 O_7$ compound the critical point can become quantum critical\footnote{Note that
in these systems there is no spontaneous symmetry breaking.} by appropriately
tuning the magnetic field \cite{Grigera} (for a related holographic study see \cite{D'Hoker:2010rz}).
Another interesting feature we notice from Figure \ref{fig:BlackHolePhaseDiagramMagnetization}
is that in all the curves with $B/\mu^2> \left(B/\mu^2\right)_I$ the magnetization saturates to a nearly constant value at low temperatures.
We emphasize that the saturation lines correspond to the cases for which the dominant $T=0$ geometry is hyperscaling violating.
Such low-temperature plateaus also occur in systems which exhibit metamagnetic phase transitions (see e.g. \cite{Krey}).
It would be interesting to make the connection to real metamagnetic materials more concrete.

When the magnetic field becomes larger than $B_{max}$, the domain walls with
 AdS$_2 \times \mathbb{R}^2$ in the deep IR cease to exist,
and the only black hole branch which can be cooled down to zero temperature always exhibits hyperscaling violation in the IR.
The magnetization suffers a discontinuous jump until we reach a critical value of the magnetic field,
$B_c >B_{max}$ at which we find only one black hole solution and
the magnetization starts changing in a continuous matter.
Thus, the \emph{line} of first order phase transitions stops at a critical point, where the phase transition becomes continuous.
This is visible in the black line shown in Figure \ref{fig:BlackHolePhaseDiagramMagnetization}, which
corresponds to the critical point $B=B_c$ and therefore to the black hole branch exhibiting hyperscaling violation at $T=0$.
Following the behavior of $m$ along the black line as the temperature increases,
we note that there is a critical temperature at which the derivative of $m/\mu$ with respect to $T/\mu$ is infinite, i.e. the phase transition is
second order (or higher).
Finally, when $B>B_c$ we no longer have a phase transition as we vary $T/\mu$.
Before we conclude this discussion we would like to point out that,
unlike in the special case $\omega=\pi/4$,
the curves we have displayed are not anti-symmetric as $B\rightarrow-B$ (remember that when $\omega=\pi/4$
\eqref{discrete sym 1} and \eqref{discrete sym 1}  combine to become a symmetry).

To summarize the results of this section,
we have seen three distinct regimes, depending on the strength of the magnetic field:
\begin{enumerate}
\item
For $0 <  \frac{B}{\mu^2}  \leq \left(  \frac{B}{\mu^2}  \right)_I$ the thermodynamically preferred black holes
can always be cooled down to domain-wall solutions with a dyonic
AdS$_2 \times \mathbb{R}^2$ geometry in the IR.
The magnetization associated with such black holes changes smoothly and monotonically as a function of temperature.
\item
When $ \left( \frac{B}{\mu^2} \right)_{I} < \frac{B}{\mu^2} \leq \left( \frac{B}{\mu^2} \right)_{max} $
the black holes favored at low temperatures approach a hyperscaling violating solution in the IR, as they are cooled to $T=0$.
On the other hand the black holes which are favored at higher temperatures
are those whose domain-walls have an AdS$_2 \times \mathbb{R}^2 $
IR description.
The latter domain-wall solutions no longer exist when $\frac{B}{\mu^2} > \left( \frac{B}{\mu^2} \right)_{max}$.
The magnetization undergoes a first order metamagnetic phase transition as a function of temperature when
$\left( \frac{B}{\mu^2} \right)_I < \frac{B}{\mu^2} < \left( \frac{B}{\mu^2} \right)_c$,
where the ratio $ \left( \frac{B}{\mu^2} \right)_c > \left( \frac{B}{\mu^2} \right)_{max}$ denotes the value at which the
transition becomes continuous.
\item
Above $\left( \frac{B}{\mu^2} \right)_c $ there is only
one black hole branch, whose $T\rightarrow 0$ limit approaches a hyperscaling violating solution
in the IR. The magnetization no longer undergoes a phase transition
as we vary the temperature of the system.
\end{enumerate}
The behavior we have discussed is generic, independently of the of the $\omega$-deformation parameter, and
its main features agree with the $\omega=\pi/4$ case studied in \cite{Gauntlett:2009bh}.

\section{Domain wall solutions}
\label{DomainWalls}

The zero-temperature description of the black holes we constructed in Section \ref{BlackHoles}
should be domain-wall geometries with AdS$_4$ asymptotics, approaching either dyonic AdS$_2 \times \mathbb{R}^2$ solutions
or hyperscaling violating geometries in the IR, depending on the value of the magnetic field.
In this section we will focus on the former case, and construct numerically domain-walls with an IR AdS$_2 \times \mathbb{R}^2$ description
by cooling down the corresponding black holes to very low temperatures\footnote{We have decreased the temperature to $T/\mu \sim 10^{-5}$
and have checked that our IR expansion is consistent with that expected for the AdS$_2 \times \mathbb{R}^2$ background
we discussed in Section \ref{IRsection}}.
Our main goal here is to discuss briefly the phase space of the solutions,
and the dependence of the free energy and magnetization on the magnetic field $B$ in the system.
We are particularly interested in features that may be entirely due to the $\omega$-deformation.

\begin{figure}[h!]
\begin{center}
\includegraphics[width=0.48\textwidth]{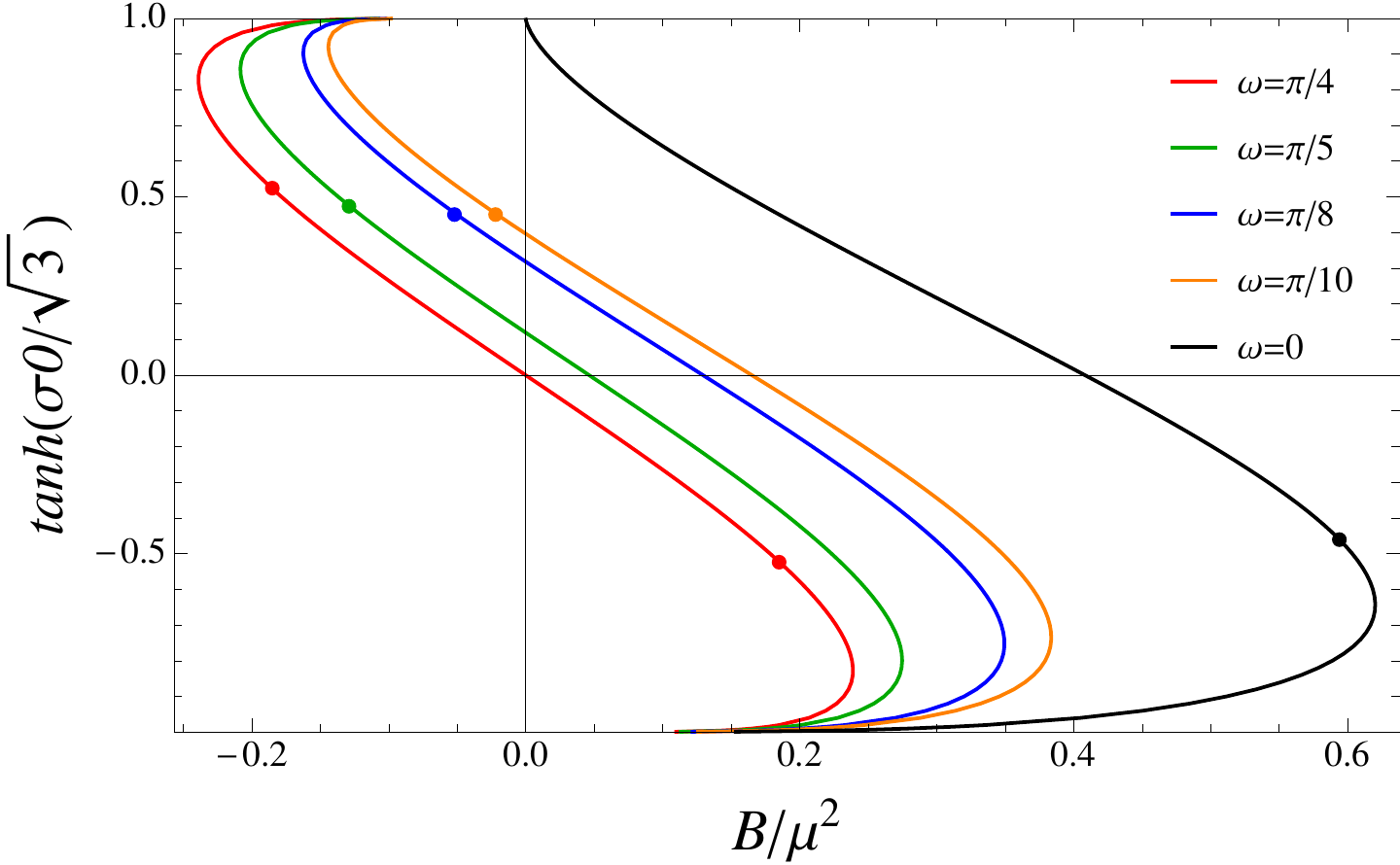}
\includegraphics[width=0.48\textwidth]{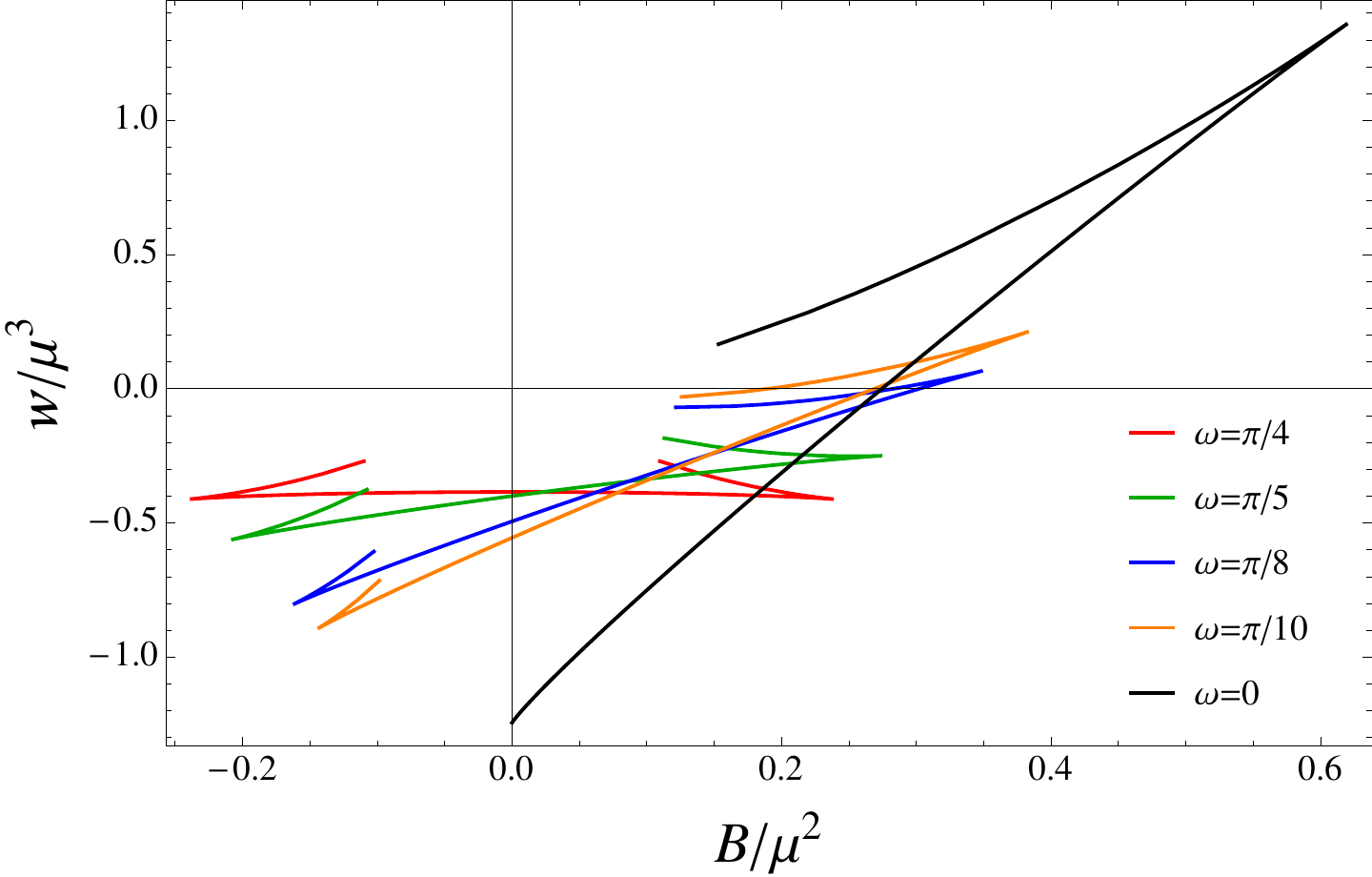}
\end{center}
\caption{Domain-wall solutions belonging to the electric family for various choices of $\omega$.
Left panel: dependence of the horizon value of the scalar on the magnetic field. The dots denote the
appearance of tachyonic modes which violate the AdS$_2$ BF found.
Right panel: free energy as a function of magnetic field.
The branches that are thermodynamically preferred are the ones which extend to $B=0$.
\label{domain wall E}}
\end{figure}

\begin{figure}[h!]
\begin{center}
\includegraphics[width=0.49\textwidth]{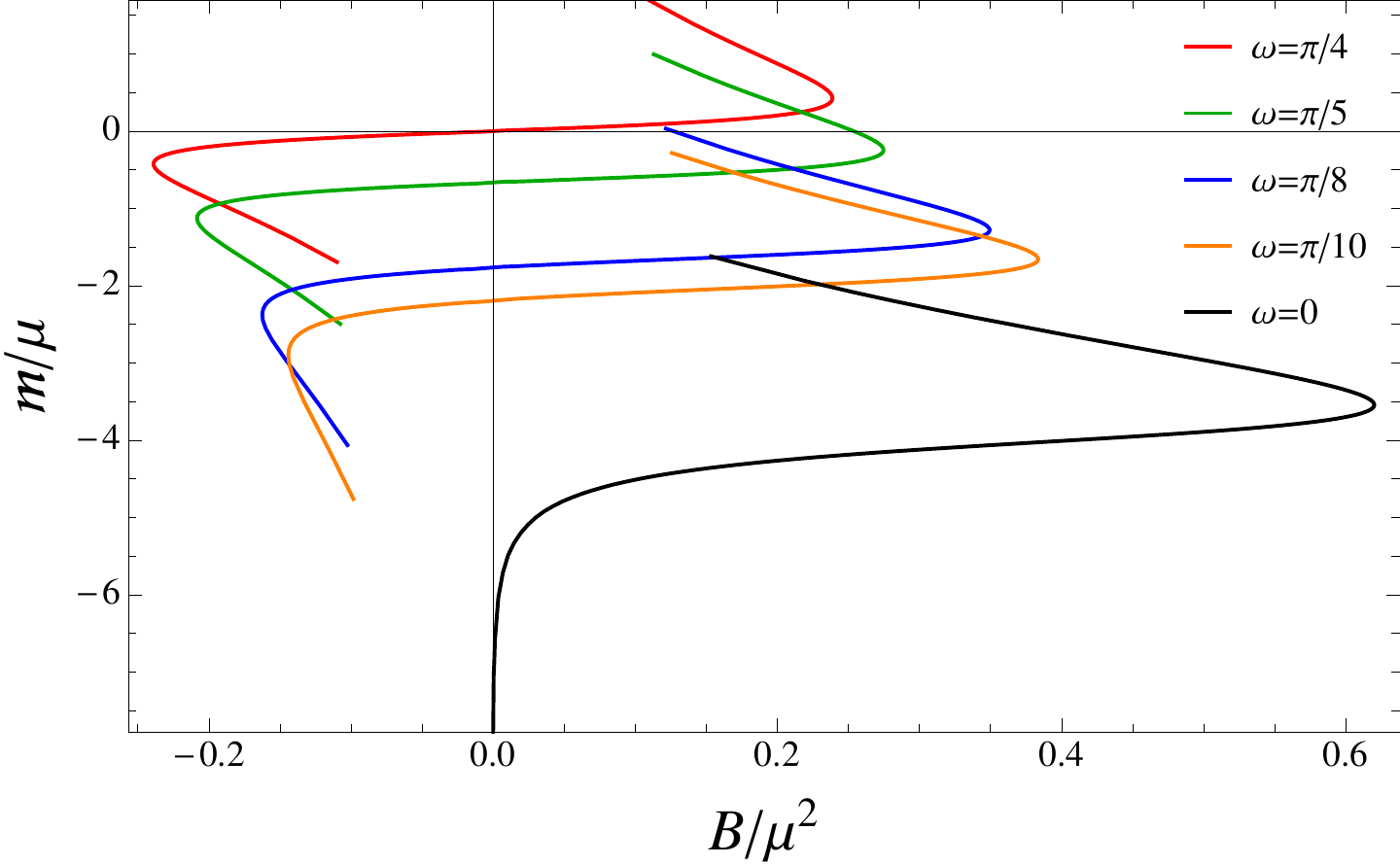}
\includegraphics[width=0.49\textwidth]{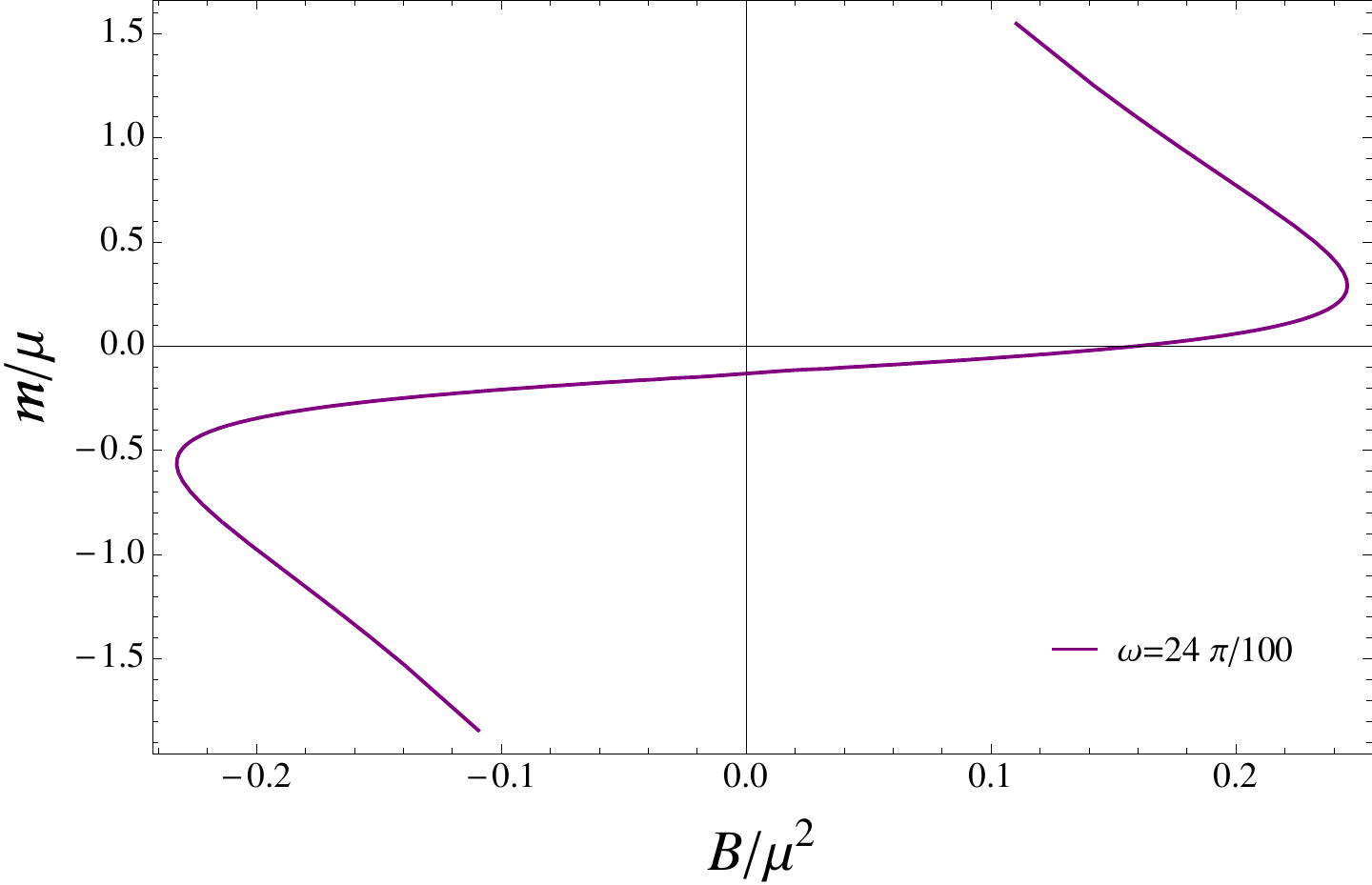}
\end{center}
\caption{Magnetic field dependence of the magnetization for the electric family of solutions, for different choices of $\omega$.
Left panel: when $B>0$ on the thermodynamically preferred branches we have $m>0$ when $\omega =\pi/4$,
and $m<0$ for the other choices of $\omega$-deformations. Right panel: the $\omega$-deformation is chosen so that
the magnetization changes sign on the thermodynamically preferred branch when $B>0$.
\label{magnetization domain wall E}}
\end{figure}

We start by discussing solutions which belong to the electric family we introduced in Section \ref{IRsection},
displayed in Figures \ref{domain wall E} and \ref{magnetization domain wall E}.
The left panel of Figure \ref{domain wall E} shows the magnetic field dependence of the IR value $\sigma_0$
of the scalar field, for several choices of $\omega$-deformation.
The red (left-most) line corresponds to the $\omega=\pi/4$ case studied in \cite{Donos:2012yu}, and is symmetric
under (\ref{additionalsymmetry}).
As the deformation parameter $\omega$ is lowered towards $\omega=0$ the solutions shift to the right
and the symmetry (\ref{additionalsymmetry}) is lost.
The dots denote the appearance of tachyonic fluctuations which violate the BF bound for AdS$_2$ and are responsible
for triggering superfluid instabilities, as we will see in detail in Section \ref{superfluid instability}.

The right panel of Figure \ref{domain wall E} shows the dependence of the free energy on the magnetic field
for each family of solutions.
Each line (corresponding to a distinct value of $\omega$) has two branches, with the thermodynamically favored one
extending all the way to $B=0$ for each choice of deformation parameter.
It is visible from this figure that the symmetry between positive and negative values of the magnetic field
is lost once we move away from the special deformation $\omega = \pi/4$.

Finally, Figure \ref{magnetization domain wall E} displays the behavior of the magnetization
as the magnetic field changes, for each family of solutions.
Let us discuss the left panel first, and for simplicity restrict our attention to the $B>0$ sector.
The red (top) line is the $\omega = \pi/4$ case, for which the magnetization is always positive and aligned with the magnetic field,
hence the system displays \emph{paramagnetism}.
In the remaining lines we find $m<0$ on the thermodynamically preferred branches,
i.e. the magnetization opposes the magnetic field and the system is \emph{diamagnetic}.
Moreover, notice that for all the cases for which $\omega \neq \pi/4$, we see a residual magnetization at zero magnetic field.

Interestingly, for certain values of the deformation parameter it is possible to find thermodynamically preferred branches
on which the magnetization can change sign, even though $B$ does not.
This is shown in the right panel of Figure \ref{magnetization domain wall E}
for the specific choice $\omega = 24 \pi/100$.
There $m$ changes sign, starting out positive at the maximum value of $B$ and becoming smaller and eventually negative as $B$ decreases
towards zero.
Thus, the system is paramagnetic for large values of the magnetic field, and becomes diamagnetic as $B$ is tuned to smaller values.
We emphasize that this behavior is not possible in the $\omega= \pi/4$ truncation.
Here we also see a residual magnetization when $B=0$.

\begin{figure}[h!]
\begin{center}
\includegraphics[width=0.60\textwidth]{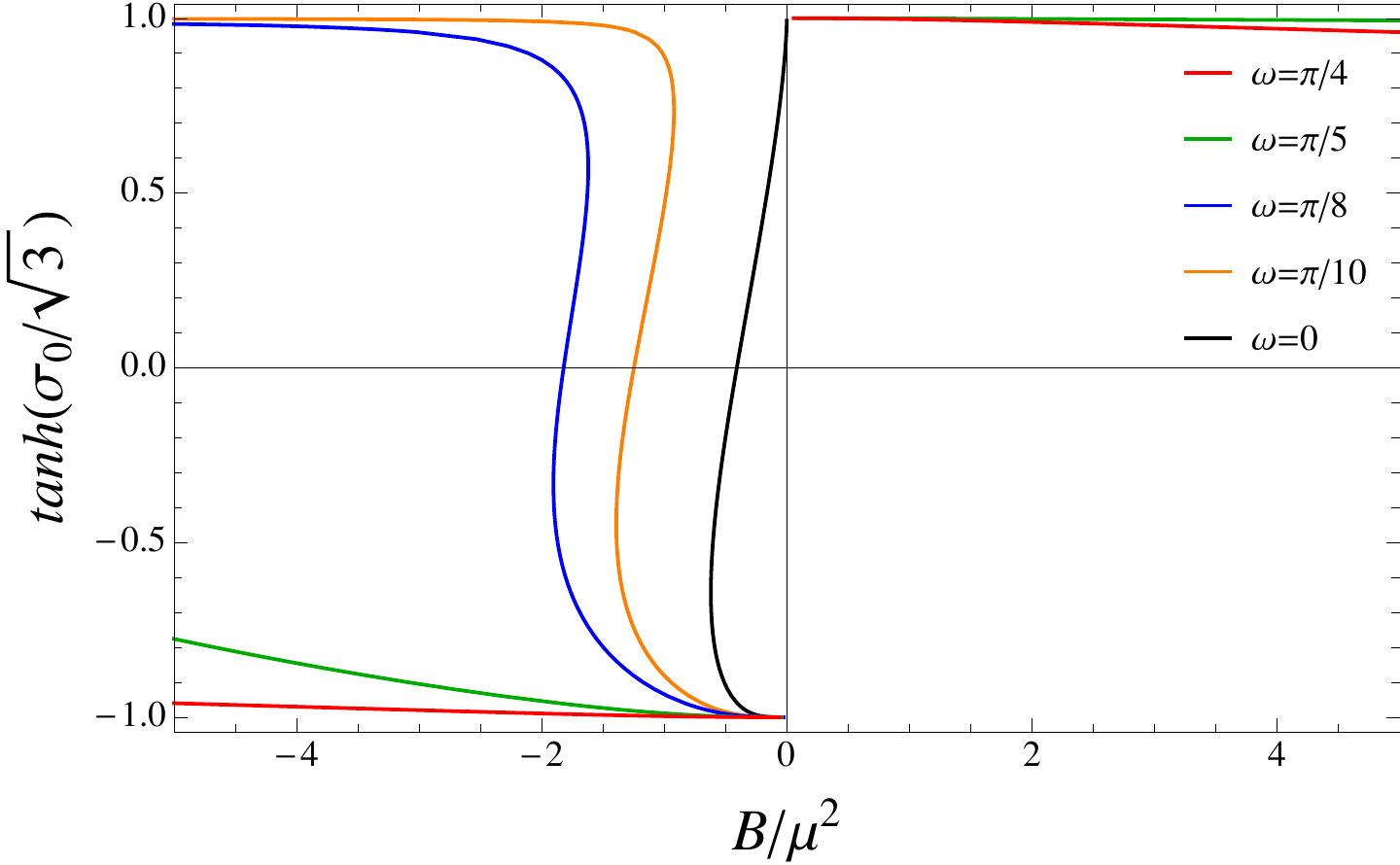}
\end{center}
\caption{Domain wall solutions belonging to the magnetic family, for different values of $\omega$
and for $g=2$\label{domainwallMsigma0}. }
\end{figure}

\begin{figure}[h!]
\begin{center}
\includegraphics[width=0.60\textwidth]{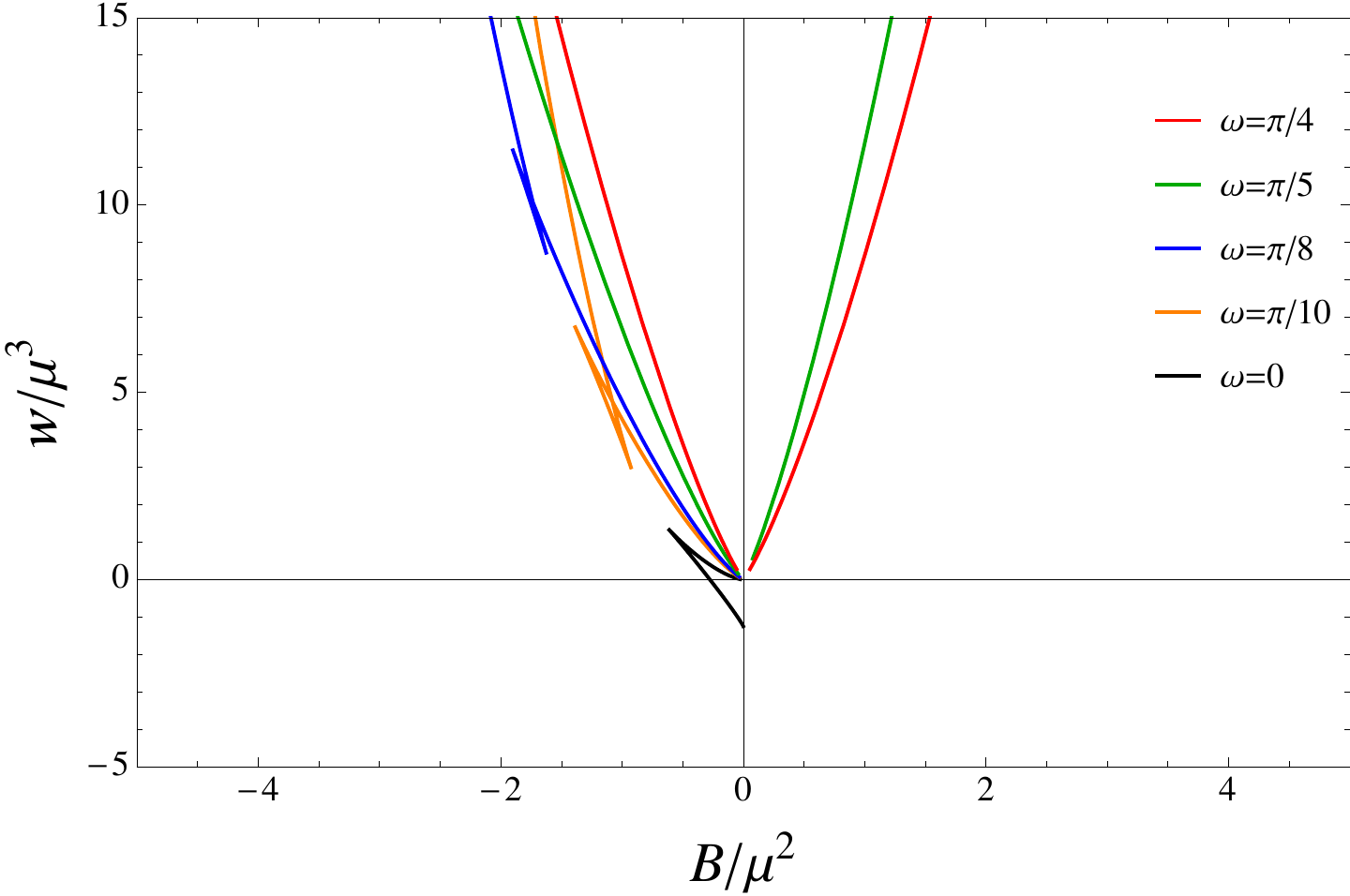}
\end{center}
\caption{Free energy of the domain-wall solutions belonging to the magnetic family, for different values of
$\omega$ $g=2$.\label{domainwallMFree}}
\end{figure}

\begin{figure}[h!]
\begin{center}
\includegraphics[width=0.60\textwidth]{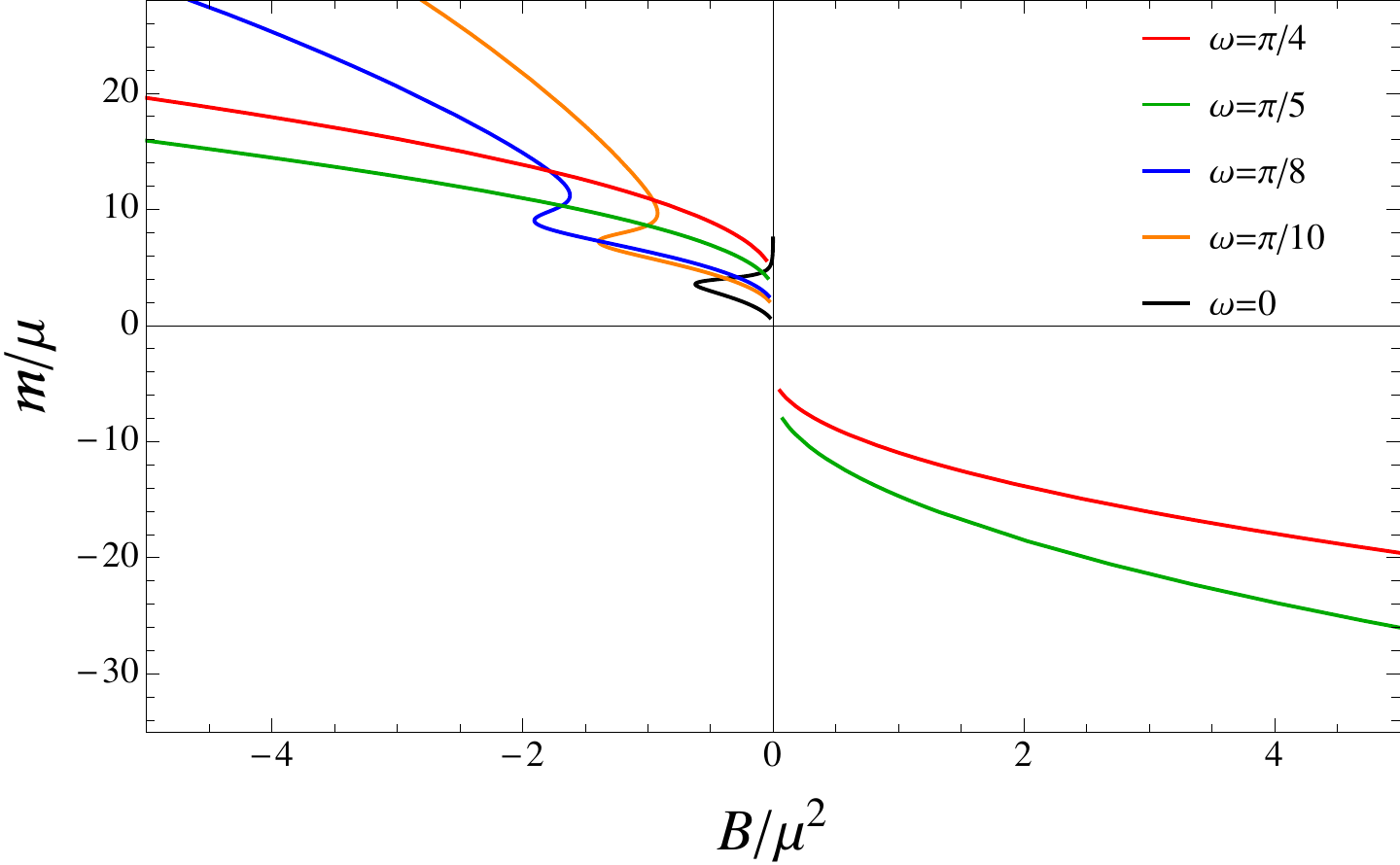}
\end{center}
\caption{Magnetization as a function of magnetic field for the the domain-wall solutions
belonging to the magnetic family, for different values of
$\omega$ $g=2$.\label{domainwallMmagnetization}}
\end{figure}

We now switch to discussing the domain-wall solutions which belong to the magnetic family, which we display in
Figures \ref{domainwallMsigma0}, \ref{domainwallMFree} and \ref{domainwallMmagnetization}.
We would like to revisit the expectation from \cite{Donos:2012yu} that solutions in the electric family would always be thermodynamically preferred,
compared to those in the magnetic family. We will provide some argument supporting this expectation, even in the case of
general $\omega$-deformations.
As before, we have constructed the domain-wall solutions numerically by cooling the black holes of Section \ref{BlackHoles}
to low temperatures, reaching $T/\mu \sim 10^{-5}$.
Figure \ref{domainwallMsigma0} displays the dependence of $\sigma_0$ on the magnetic field\footnote{In this figure
we have made use of the symmetry (\ref{discrete sym 3})
to ensure that $T/\mu$ is always positive.}.
We find that when $\tanh(\sigma_0/\sqrt{3})$ is very close to one it is difficult to construct the solutions numerically.
Thus, we were not able to obtain the $\omega =\pi/8, \pi/10$ branches which should appear in the first quadrant.
Finally, Figures \ref{domainwallMFree} and \ref{domainwallMmagnetization} show, respectively, the free energy and magnetization
as a function of magnetic field.
We were able to probe only a small region where both electric and magnetic families coexist.
In this region, the domain-walls in the electric family are always thermodynamically preferred.
Moreover, as we heat up the solutions, the free energy of the black holes coming from the magnetic (electric)
family increases (decreases). This provides further evidence which indicates
that the electric branch is always thermodynamically favored.
Clearly it would be valuable to confirm this with further studies.

\section{Superfluid instability}
\label{superfluid instability}

So far we have restricted our attention to solutions for which the charged complex scalar field vanishes, $\rho \, e^{i\chi}= 0$.
However, when the background geometry is described by black holes whose
extremal limit has a near-horizon AdS$_2 \times \mathbb{R}^2$,
we expect the field to condense at some critical temperature $T_c$, spontaneously breaking the $U(1)$ and
triggering a superfluid phase transition \cite{Gubser:2008px,Hartnoll:2008vx,Hartnoll:2008kx}.
In this section we will examine the appearance of such superfluid instabilities in the dyonic black hole backgrounds
we constructed in Section \ref{BlackHoles}, when the magnetic field is in the range (\ref{Brange}).
However, the presence of unstable modes can already be anticipated
by considering linearized fluctuations $\delta\rho$ of the charged scalar about the AdS$_2 \times \mathbb{R}^2$
solutions\footnote{We will only consider solutions belonging to the electric family, since they should be thermodynamically favored.} of Section \ref{IRsection}, for which $\rho=0$.
The system will be unstable when there are tachyonic modes which violate the BF bound for AdS$_2$.
To see this more explicitly, we examine the equation of motion for the scalar perturbation\footnote{We can set the
phase to $\chi=0$ by a gauge choice.} $\delta\rho$ on this background, which takes the form
\be
\square \delta\rho - 4 g^2 A^2\, \delta\rho -\fft{\del^2 V}{\del\rho^2}\, \delta\rho =0\,,
\ee
and hence
\be
\square\delta\rho = g^2 \left[ B^2 (x_1^2+x_2^2) - 4 \ell^2 E^2
+  \left(\cosh(\sqrt3 \sigma_0) -  3\cosh\fft{\sigma_0}{\sqrt3}  +
   4\cos2\omega\, \sinh^3\fft{\sigma_0}{\sqrt3} \right)  \right] \delta\rho \,. \nn
\ee
We take the fluctuation to describe the first Landau level, which we expect to condense first \cite{Hartnoll:2008kx},
\be
\delta\rho= u(t,r)\, e^{-\ft12 g |B|(x_1^2 + x_2^2)}\, .
\ee
It then follows that
\be
\square_{AdS_2}\, u = M^2\, u \,,
\ee
where $\square_{AdS_2}$ is the d'Alembertian on the AdS$_2$ spacetime and
\be
M^2 = 2 g \, |B| - 4 g^2\ell^2\, E^2  - 3g^2 \cosh\fft{\sigma_0}{\sqrt3}
 + g^2 \cosh(\sqrt3 \sigma_0) +
   4 g^2 \cos2\omega\, \sinh^3\fft{\sigma_0}{\sqrt3} \,.
\ee
Superfluid instabilities are triggered when the mass of the fluctuation becomes smaller than the
BF bound for AdS$_2$, \emph{i.e.} when $M^2 < m_{BF}^2 = -g^2/4$, which in our case is
\beq
\label{BFcondition}
\boxed{
 2 g \, |B| - 4 g^2\ell^2\, E^2  - 3g^2 \cosh\fft{\sigma_0}{\sqrt3}
 + g^2 \cosh(\sqrt3 \sigma_0) +
   4 g^2 \cos2\omega\, \sinh^3\fft{\sigma_0}{\sqrt3} < - \frac{g^2}{4} \, . \; }
\eq
Notice that stronger (weaker) electric (magnetic) fields enhance the instability window.
\begin{figure}[h!]
\begin{center}
\includegraphics[width=0.60\textwidth]{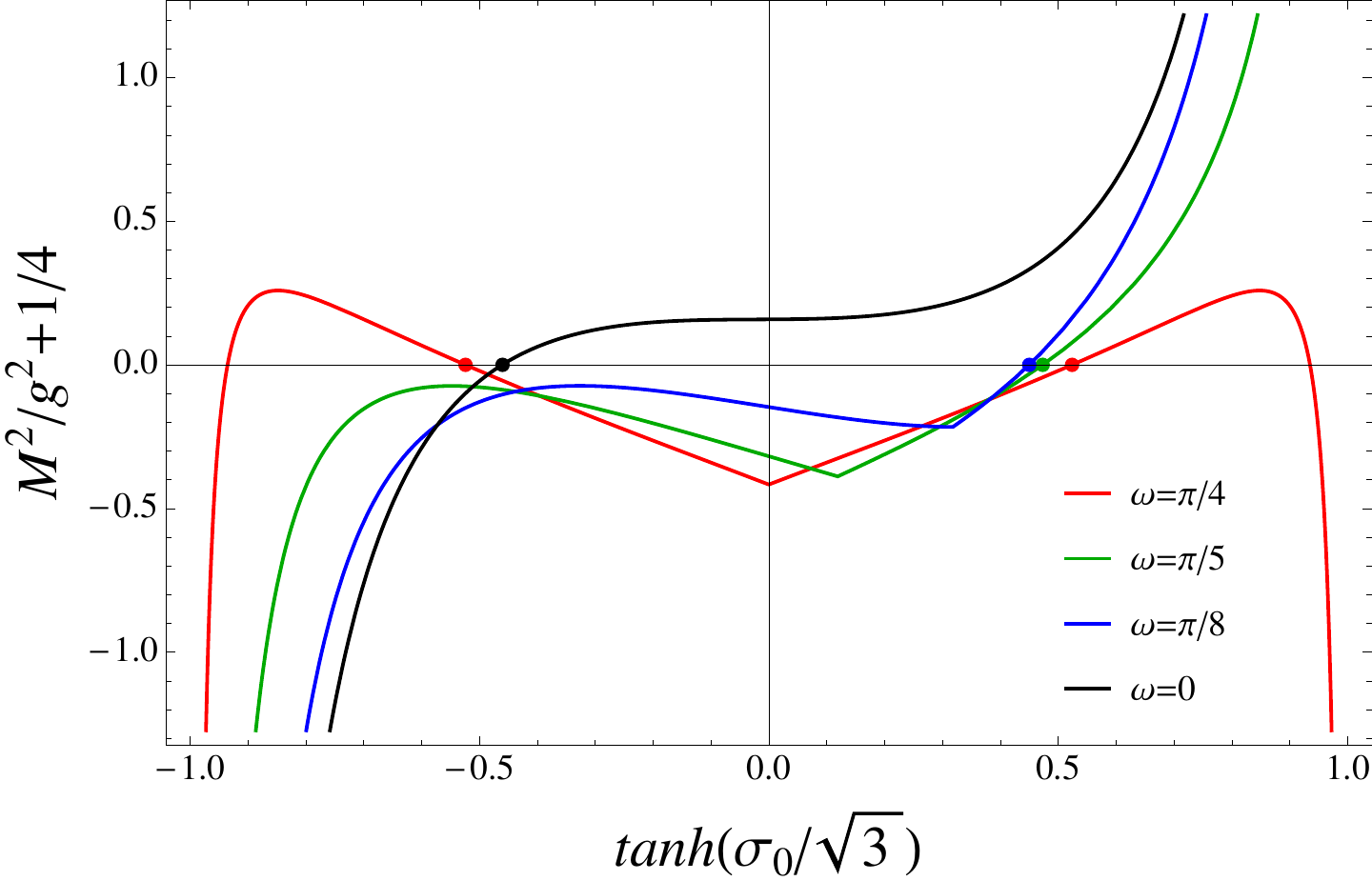}
\end{center}
\caption{Dependence of $M^2/g^2+1/4$ on $\tanh(\sigma_0/\sqrt{3})$ for the domain-wall solutions shown in Figure \ref{domain wall E},
for different values of the $\omega$-deformation and with $g=2$.
When the curves cross zero and become negative, the mass of the charged scalar fluctuation violates the AdS$_2$ BF bound,
causing the zero temperature domain-wall to be unstable. The dots denote the point at which the instability sets in.
\label{fig:BFboundviolation}}
\end{figure}

In Figure \ref{fig:BFboundviolation} we show the mass of the fluctuation $\delta\rho$
(more precisely, the shifted mass $M^2/g^2 + 1/4$) in the AdS$_2 \times \mathbb{R}^2$ background
as a function of the IR value $\sigma_0$ of the scalar, as given in (\ref{BFcondition}).
Different curves corresponds to different choices of $\omega$-deformation.
Superfluid instabilities are expected when each curve becomes negative, corresponding to the fluctuation violating
the BF bound for AdS$_2$.
The onset of the instability is denoted by a dot in the figure, and occurs when a line
crosses the horizontal axis.
Inspecting Figures \ref{domain wall E} and \ref{fig:BFboundviolation}, we note that  when
$\tanh(\sigma_0/\sqrt{3})<0$ (which corresponds to $B/\mu^2>0$)
the BF bound is \emph{always} violated for the curves with $\omega= \pi/5, \pi/8, \pi/10$.
We will return to this point shortly.

What we are really interested in, however, are superfluid instabilities appearing in the dyonic black holes
we studied in Section \ref{BlackHoles}, with an IR AdS$_2 \times \mathbb{R}^2$ zero temperature description.
Recall that the latter are always thermodynamically preferred when the field is relatively low, and in the range (\ref{Brange}).
Thus, our instability analysis will only describe the regime $B/\mu^2 \leq \left(B/\mu^2\right)_I$.
To determine whether the charged scalar field can condense at non-zero temperature, we will ask whether a normalizable mode
of $\rho$ appears at some critical temperature denoted by $T_c$, for an appropriate range of magnetic field.
Since we are interested in breaking the abelian gauge symmetry spontaneously, we do not want to generate a non-normalizable
mode for $\rho$, corresponding to a source for the dual operator $\cal{O}_\rho$.
To this end, we want to solve for the linearized fluctuation $\delta\rho$
assuming a background of the form of (\ref{ansatz}).
We take the fluctuation to be of the form
\be\label{R equ}
\delta\rho= R(r)\, e^{-\ft12 g |B|(x_1^2 + x_2^2)} \, .
\ee
Substituting it into the equation of motion for $\rho$ and working to linear order, we find
\be
r^{-2}e^{\frac{\beta (r)}{2}} \left(r^2 f(r) e^{-\frac{\beta (r)}{2}} R'(r)\right)'-\left(\frac{2 g \left| B\right|}{r^2}+G(\sigma)-\frac{4 g^2 e^{\beta (r)} \phi (r)^2}{f(r)}\right)R(r)=0\\
\ee
where
\be
G(\sigma)=g^2 \left(-4 \cos (2 \omega ) \sinh ^3\left(\frac{\sigma (r)}{\sqrt{3}}\right)-3 \cosh \left(\frac{\sigma (r)}{\sqrt{3}}\right)+\cosh \left(\sqrt{3} \sigma (r)\right)\right)\nonumber.
\ee
At the horizon, the radial perturbation $R(r)$ has an expansion of the form
\be
R(r)=c_1+c_2\log(r-r_+) + \ldots \; .
\ee
We set $c_2=0$ to ensure that $R(r)$ is regular .
On the other hand, the boundary behavior of $R(r)$ is given by
\be
R(r)=\frac{R_1}{r}+\frac{R_2}{r^2} \, ,
\ee
with $R_1$ and $R_2$ denoting, respectively, the source and the VEV of the operator dual to the charged scalar.

By varying $T/\mu$, we find a solution for the $\delta \rho$ perturbation which has $R_1=0$ and $R_2 \neq 0$,
indicating that the field indeed condenses and the symmetry breaking is spontaneous, as desired.
\begin{figure}[h!]
\begin{center}
\includegraphics[width=0.60\textwidth]{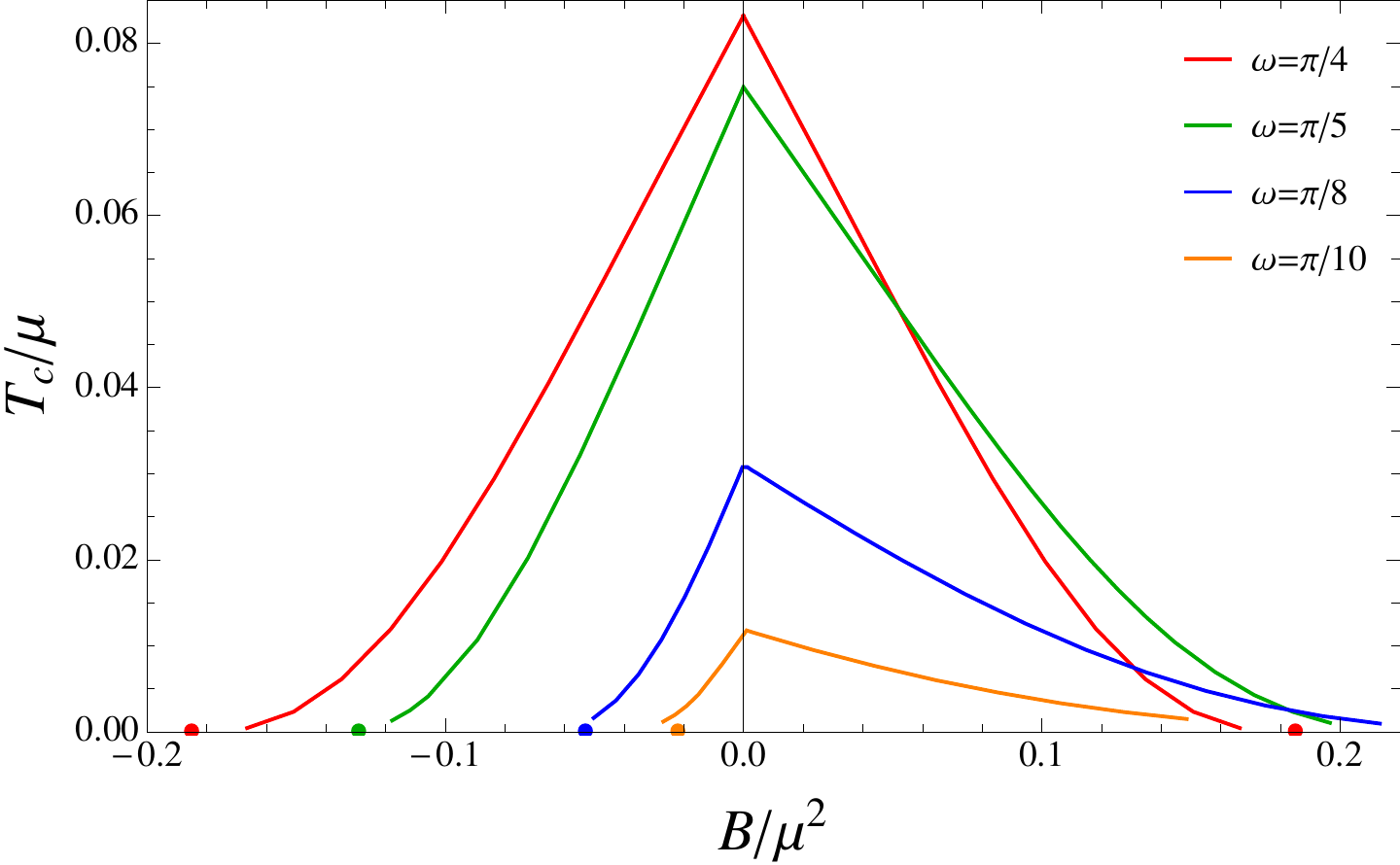}
\end{center}
\caption{Critical temperature at which the superfluid phase transition sets in, for different choices of $\omega$ and for $g=2$.
The $\omega$ parameter for each curve decreases from top to bottom.
The dots on the horizontal axis denote the points at which the mass of the charged scalar field fluctuation
about the $T=0$ AdS$_2 \times \mathbb{R}^2$ IR geometry violates the AdS$_2$ BF bound, as shown in Figure \ref{fig:BFboundviolation}.
\label{fig:CriticalTemperature}}
\end{figure}
Our results are shown in Figure \ref{fig:CriticalTemperature}, where we plot the critical temperature $T_c$
at which the field starts to condense as a function of the external magnetic field, for different choices of $\omega$.
We would like to highlight a few features of this analysis.
First, as $T_c \rightarrow 0 $ we expect the curves to approach the dots denoting superfluid instabilities
of the zero temperature AdS$_2 \times \mathbb{R}^2$ geometry.
Notice that there is a slight deviation between the dots and the low-$T_c$ regime of some of the curves.
We expect this discrepancy to be a reflection of the fact that we are working in a linearized approximation,
and that a fully back-reacted analysis would resolve it.
In fact, recall that in \cite{Hartnoll:2008kx}, taking into account back-reaction lead to a suppression of $T_c$
compared to the probe limit result. The disagreement between the two cases became more important as $T_c \rightarrow 0$,
\emph{i.e.} away from the regime of validity of the probe approximation.

An interesting feature is that as we move $\omega$ away from $\pi/4$ while keeping $|B|$ very small, $T_c$ is suppressed (the
suppression is not always present in the regime where the magnetic field is large).
Thus, as $\omega$ decreases one has to reach lower temperatures in order to access the superfluid phase.
We expect this to be especially a factor for the $\omega=0$ case, for which we were not able to reach $T_c$ in our numerics.

When $B<0$, we can easily see from Figure \ref{fig:CriticalTemperature} that there is a range of magnetic field
for which a condensate does not form. In particular, if the magnitude of the field is too large, it will
prevent the formation of a superfluid phase, consistent with expectations from the Meissner effect.
On the other hand, when $B>0$ as we decrease $\omega$ (moving from top to bottom in the figure)
the range of $B$ which allows for a superfluid instability is seemingly becoming larger.
In fact, the curves seem to flatten out as $B/\mu^2$ increases.
This is consistent with the structure of Figure \ref{fig:BFboundviolation}, where we saw that the blue and green
curves always violated the AdS$_2$ BF bound for $B>0$.
Thus, the superconducting phase naively seems to survive even in very strong magnetic fields.
However, we should keep in mind that our instability analysis assumes that the magnetic field is in the range (\ref{Brange}),
and breaks down when $B/\mu^2 = \left(B/\mu^2\right)_I$, at which point the thermodynamically preferred background is hyperscaling violating.
Let's return briefly to the behavior of the $\omega=0$ curve.
By combining Figures \ref{domain wall E} and \ref{fig:BFboundviolation} we see that, when $\omega$ is nearly zero, the value of $B$ for which
the fluctuation $\delta \rho$ is tachyonic is already very close to $B_{max}$.
Thus, there is a very narrow window in which the superfluid instability can occur.
Moreover, this corresponds to a very low $T_c$, making it even harder to observe numerically.

\subsection{Competition with stripe instabilities}

As we already stressed in the introduction, AdS$_2 \times \mathbb{R}^2$ solutions
are also known to be unstable to the formation of spatially modulated phases,
triggered by modes which violate the AdS$_2$ BF bound and break translational symmetry
(see the analysis of \cite{Donos:2012yu} for the dyonic case).
Thus, we expect the domain-walls we constructed in Section \ref{DomainWalls} -- which have an AdS$_2$ factor in their IR
description -- to also suffer from striped instabilities.
By the same token, there should be spatially modulated tachyonic modes in the
non-zero temperature generalizations of these
solutions -- the dyonic black holes of Section \ref{BlackHoles} -- provided\footnote{In this regime the thermodynamically preferred
black holes \emph{always} approach AdS$_2 \times \mathbb{R}^2$ in the IR as $T \rightarrow 0$.} the magnetic field
is in the range (\ref{Brange}).

The authors of \cite{Donos:2012yu} discussed the competition between superfluid and spatially modulated instabilities
in the context of the $\omega = \pi/4$ theory.
In particular, for the class of dyonic AdS$_2 \times \mathbb{R}^2$ solutions to the theory they found that
the existence of striped instabilities was independent of the value of the magnetic field.
As a result, increasing the magnetic field should act to suppress the superfluid instabilities compared to the striped ones.
In particular, \emph{a priori} at non-zero temperature one expects both types of instabilities to be
generically present\footnote{Clearly one should also determine which instability is triggered first,
by comparing their critical temperatures at a given vale of $B$.}
as long as the field is in the range
$0<|B| <|B^{SC}_c| < |B_I|$,
where $B^{SC}_c$ denotes the point at which the charged scalar no longer condenses (i.e. the critical temperature for the
superfluid phase transition becomes $T_c=0$).
On the other hand, when $|B^{SC}_c| < |B| < |B_I|$ the superfluid phase is no longer accessible and
only striped instabilities should survive.

For the $\omega$-deformed case the analysis of spatially modulated perturbations of the dyonic AdS$_2 \times \mathbb{R}^2$
solutions is analogous to that of \cite{Donos:2012yu}.
In particular, the spectrum of the scaling dimensions for the fluctuations is the same,
since a duality rotation relates the solutions of the
$\omega$-deformed theory to those of \cite{Donos:2012yu}, when the charged scalar field is turned off.
The main difference in the $\omega$-deformed case comes from the behavior
at non-zero temperature, and is
due to the fact that the theory is no longer invariant under $B \rightarrow - B$.
In particular, recall from Figure \ref{fig:CriticalTemperature} that when $B>0$ the charged scalar
naively appears to condense for arbitrarily high values of $B$, in contrast with expectations from the Meissner effect.
However, since the instability analysis is only valid up to $B=B_I$, the latter value
sets a natural cutoff for the existence of a superfluid phase.
Still, for the $\omega$-deformed theories we expect to have \emph{both} stripe and superfluid phases in the entire range
$0<B<B_I$, unlike for the $\omega=\pi/4$ theory.
On the other hand the behavior when $B<0$ is analogous to that of \cite{Donos:2012yu}, with both instabilities present
when $-|B^{S.C.}_c| < B < 0 $ and striped ones alone in $ -|B_I| < B < -|B^{S.C.}_c|$.

It would be useful to determine the temperature at which the spatially modulated instabilities are triggered,
and in particular whether it is above or below the one associated with the onset of the superfluid instability.
Answering this question would be a first step towards better understanding the ultimate ground states of the $\omega$-deformed
theories. Clearly, the question of back-reaction on the geometry is even more important, although challenging to investigate.
We should also mention that the hyperscaling violating geometries themselves are believed to suffer,
in certain cases, from striped instabilities.
For discussions of this question we refer the reader to e.g. \cite{Cremonini:2012ir,Iizuka:2013ag,Cremonini:2013epa}.
Thus, it is possible that we would find spatially modulated tachyonic modes even when the zero temperature geometry in the deep IR
exhibits hyperscaling violation, for $|B|>|B_I|$.

\section{Conclusions}

The $\omega$-deformed supergravity truncations we have studied in this paper
admit a rich variety of phases,
which can be accessed by appropriately tuning the magnetic field of the system and varying its temperature.
An interesting structure already emerges when we consider truncations which retain, in addition to a $U(1)$ gauge field,
a single neutral scalar. Dyonic black hole solutions in this case
exhibit a line of first order metamagnetic phase transitions -- describing a sudden change in the magnetization --
once $B$ is sufficiently strong.
Moreover, as they are cooled down to zero temperature, they behave either as a diamagnetic or a paramagnetic material,
depending again on the strength of $B$ and the particular choice of $\omega$.
In these truncations the deep IR region of the extremal geometries is described by either
dyonic AdS$_2 \times \mathbb{R}^2$ or a
solution with a non-trivial dynamical critical exponent and hyperscaling violation. It is precisely the tension
between black hole branches with these different IR descriptions which is responsible for the metamagnetic phase transition.

In less restrictive truncations the presence of a complex scalar charged under the $U(1)$
allows for the existence of low-temperature superconducting phases, which are expected in models of this type
when the magnetic field is not too large.
However, in the $\omega$-deformed theories the mechanism by which the superconducting instability ceases to exist is different
depending on whether the magnetic field is positive or negative.
In particular, when $B<0$ the charged scalar stops condensing in the black hole background
at a critical value of $B$, consistent with intuition
from the Meissner effect that a strong enough magnetic field should destroy superconductivity.
The corresponding extremal near-horizon AdS$_2 \times \mathbb{R}^2$ geometries
exhibit superfluid instabilities
only within a certain range for $B$, which is typically \emph{smaller} than the range in which the AdS$_2$ solution
exists.
This behavior is visible in Figures \ref{fig:BFboundviolation} and
\ref{fig:CriticalTemperature}, and was also observed in
the $\omega=\pi/4$ truncation studied in \cite{Donos:2012yu}.
When $B>0$, however, the mechanism that
halts the superconducting phase is different.
As long as the value of $\omega$ is not too close to $\pi/4$,
the tachyonic modes of the extremal IR AdS$_2 \times \mathbb{R}^2$ are present for
arbitrarily strong values of the field (see for example
the $\omega = \{\pi/5, \pi/8, 0\}$ curves in
Figures \ref{fig:BFboundviolation} and \ref{fig:CriticalTemperature}).
As a result, in these theories the superconducting phase ceases to exist only when the extremal geometry is no longer described by a
domain-wall with an IR AdS$_2 \times \mathbb{R}^2$.

This asymmetry between positive and negative values of $B$
also affects in an interesting way the interplay between superconducting and striped phases,
with the latter triggered by spatially modulated modes which violate the AdS$_2$ BF bound.
In analogy with \cite{Donos:2012yu}, in our truncation striped instabilities should be insensitive to the strength of
the magnetic field, as long as it lies within the range specified in
(\ref{Brange}).
Thus,
while for $B<0$ there will be a window in which only striped phases are present,
when $B$ is positive and within the range (\ref{Brange}) we expect to find
both classes of instabilities.
Which instability is
triggered first will of course depend on the
competition between their critical temperatures.

Clearly one of the more challenging questions associated with these
theories is the issue of back reaction.
It would be valuable to determine the fully non-linear backgrounds
associated with such phases,
to shed light on the vacuum structure of the theory.
A related question is which features, if any, are due entirely to
the presence of the $\omega$-deformation.
A hint could come from the asymmetry between $B$ and $-B$,
which affects the competition between striped and superconducting
phases and therefore the geometric properties of the ground state.
We would also like to gain a better understanding of the new set of
AdS$_4$ vacua that we identified in Appendix B.  These have instabilities
due to the occurrence of linearised scalar fluctuations that violate the
BF bound, but in one of the cases one could consider a consistently-truncated
sub-theory within which the new AdS$_4$ vacuum would be stable.
In particular, it would be interesting to construct
domain-wall geometries which interpolate between two $AdS_4$ fixed points,
and ask whether any
intermediate scaling
behavior is possible along the flow, as in the construction of \cite{Bhattacharya:2014dea}.
Finally, while some of the features we have observed in this paper have analogs in the behavior of strongly correlated
materials in the presence of a magnetic field,
we would like to refine these ideas further and make these connections more concrete.
We leave these questions to future work.

\vskip 1in
{\noindent\large  \bf Acknowledgments}
\vskip 0.1in

We are grateful to Artem Abanov, Jerome Gauntlett, Blaise Gout\'eraux, Fuxiang Li, Yun Wang and Jackson
Wu for helpful discussions.  We are especially grateful to Aristos
Donos for extended discussions.  S.C. and C.N.P. gratefully acknowledge the
hospitality of the Cambridge-Mitchell Collaboration for hospitality at
the Great Brampton House workshop during the course of this work.
Y.P. is grateful to the Mainz Institute for Theoretical Physics (MITP)
for its hospitality and its partial support during the completion of this work.
The work of C.N.P. is supported in part by DOE grant DE-FG02-13ER42020.

\appendix

\section{Duality Rotation of Physical Quantities}
\label{AppendixDuality}

In this appendix we show that when the charged scalar field is turned off,
the solutions of the $\omega$-deformed theory can be obtained from the undeformed, $\omega=0$ theory.
The latter is described by the Lagrangian
\be
e^{-1}{\cal L}^0_{F} = -U_0(\sigma)\, F^{\mu\nu} F_{\mu\nu} - W_0(\sigma)\,
F^{\mu\nu}\, {^*\! F}_{\mu\nu}\,\quad
U_0(\sigma) =
e^{\sqrt{3}\sigma},\quad
W_0(\sigma) =0.
\label{undeformedL}
\ee
If we define a 2-form $G$ through
\be
G=U_0{^*\! F}-W_0F \, ,
\ee
the equation of motion derived from (\ref{undeformedL}) and the Bianchi identity can be summarized as
\be
dG=0,\quad dF=0 \, .
\ee
As discussed in \cite{Lu:2014fpa}, the above set of equations is invariant under an $Sp(2,\mathbb{R})$ transformation.
In other words, after an $Sp(2,\mathbb{R})$ rotation,
\be
\begin{pmatrix} F_{\Lambda}\cr G_{\Lambda}\end{pmatrix} =
 \Lambda\, \begin{pmatrix} F\cr G\end{pmatrix}\,,\qquad
\Lambda\in Sp(2,\mathbb{R}),
\ee
the quantities $F_{\Lambda}$ and $G_{\Lambda}$ still satisfy
\be
dG_{\Lambda}=0,\quad dF_{\Lambda}=0 \, .
\ee
Meanwhile, $G_{\Lambda}$ can be expressed as
\be
G_{\Lambda}=U_{\Lambda}(\sigma){^*\! F}_{\Lambda}-W_{\Lambda}(\sigma)F_{\Lambda},
\ee
where $U_{\Lambda}(\sigma)$ and $W_{\Lambda}(\sigma)$ are scalar functions of $\sigma$.
In particular, if we choose the duality rotation matrix to be
in the $U(1)$ subgroup of $Sp(2,\mathbb{R})$,
\be
\Lambda= \begin{pmatrix} \cos\omega &\quad -\sin\omega\cr
                   \sin\omega & \quad\cos\omega\end{pmatrix},
\ee
the corresponding scalar functions $U_{\Lambda}(\omega)$ and $W_{\Lambda}(\omega)$ are given by
\be
U(\sigma) =
\frac1{\cosh\sqrt3\sigma -\cos2\omega\sinh\sqrt3\sigma}
\,,\qquad
W(\sigma) = \frac{\sin2\omega\sinh\sqrt3\sigma}{
\cosh\sqrt3\sigma- \cos2\omega\sinh\sqrt3\sigma}.
\ee
Therefore, $dG_{\omega}=0$ and $dF_{\omega}=0$ are just the equation of motion and Bianchi identity of
the $\omega$-deformed theory whose Lagrangian is given in Section \ref{TruncationSection}.

From the analysis above we see that one can generate a solution to the $\omega$-deformed theory
by performing a $U(1)$ rotation of a solution in the undeformed theory.
For instance, to obtain the electromagnetic fields for an AdS$_2\times \mathbb{R}^2$ geometry in the $\omega$-deformed theory
we first solve $(E, B)$ from the $\omega=0$ theory, where
$(E, B)$ satisfy
\bea
\ell^{-2} &=&- V\,,\qquad E^2+B^2 = - \fft{V}{2} e^{-\sqrt{3}\sigma_0}\,,\nn\\
0&=& 2\sqrt{3} (B^2-E^2)\, e^{\sqrt{3}\sigma_0} + V'(\sigma_0)\,.
\eea
From these relations we can easily solve for $E_0=E(\omega=0)$ and $B_0=B(\omega=0)$ in terms of $\sigma_0$.
The solution in the $\omega$-deformed theory can then be obtained via
\be
E_{\omega}=\cos\omega E_0-\sin\omega e^{\sqrt{3}\sigma_0} B_0,\quad B_{\omega}=\cos\omega B_0+\sin\omega e^{\sqrt{3}\sigma_0} E_0 \, .
\ee
Modulo the overall sign change of $(E, B)$, there are two families of solutions.
To match with the convention used in \cite{Donos:2012yu}, we choose the solution in the electric family to be,
\bea
E_{0}^{(e)}/g&=&\sqrt{\frac{(3-\tanh\frac{\sigma_0}{\sqrt{3}})(1-\tanh\frac{\sigma_0}{\sqrt{3}})}{2(1+\tanh\frac{\sigma_0}{\sqrt{3}})^2}},\nn\\
B_{0}^{(e)}/g&=&
-\sqrt{\frac{(3+\tanh\frac{\sigma_0}{\sqrt{3}})(1-\tanh\frac{\sigma_0}{\sqrt{3}})}{2(1+\tanh\frac{\sigma_0}{\sqrt{3}})^2}};
\eea
while the one in the magnetic family is given by
\bea
E_{0}^{(m)}/g&=&
-\sqrt{\frac{(3-\tanh\frac{\sigma_0}{\sqrt{3}})(1-\tanh\frac{\sigma_0}{\sqrt{3}})}{2(1+\tanh\frac{\sigma_0}{\sqrt{3}})^2}},\nn\\
B_{0}^{(m)}/g&=&
-\sqrt{\frac{(3+\tanh\frac{\sigma_0}{\sqrt{3}})(1-\tanh\frac{\sigma_0}{\sqrt{3}})}{2(1+\tanh\frac{\sigma_0}{\sqrt{3}})^2}}.
\eea

\section{Additional AdS$_4$ Vacua}
\label{AppendixAdS4}

   In the four-scalar truncation that we we are considering in this paper,
the scalar potential $V$ for the $\omega$-deformed theory is given by eqn
(\ref{pot1}).  Note that it depends on the three scalars $(\sigma, \rho, x)$,
but is independent of the fourth scalar $\chi$.  We may seek AdS$_4$
vacua by looking for stationary points of the potential.  We first note
that the condition $\del V/\del x=0$ implies
\be
[(R^2 \cos^2\omega + \sin^2\omega)\, \cosh\rho +
  (\cos^2\omega + R^2 \sin^2\omega)\, \cosh x]\, \sinh  x =0\,.
\ee
Since the quantity in the square brackets is always positive for real values
of the fields and $\omega$ parameter, it follows that we must have
\be
x=0
\ee
at all stationary points.  Setting $x=0$, we then find from
$\del V/\del \rho=0$
that
\be
[(R^6 \cos^2\omega + \sin^2\omega)\, \cosh\rho -3 R^2\,
(R^2 \cos^2\omega + \sin^2\omega)\, \cosh\rho]\, \sinh\rho=0\,.\label{Vrho}
\ee
This then gives either the trivial stationary point
\be
\rho=0\,,\qquad R=1\,,\qquad x=0\,,\label{trivial}
\ee
(i.e. the standard AdS$_4$ vacuum that is supersymmetric in the full
${\cal N}=8$ theory),
or else the square bracket in (\ref{Vrho}) vanishes,  implying
\be
 \cosh\rho = \fft{3 R^2\, (R^2 \cos^2\omega + \sin^2\omega)}{
          (R^6 \cos^2\omega + \sin^2\omega)}\,.\label{rhosol}
\ee
We shall focus on the $\rho\ne0$ non-trivial stationary points from now on.

Inserting (\ref{rhosol}), together with $x=0$, into the potential then gives
from $\del V/\del R=0$  a factorised equation that implies either
\be
R^6\, (R^4-5)\cos^2\omega - (5R^4-1)\sin^2\omega=0\,,\label{VR1}
\ee
or else
\be
R^8 -2R^2\, (R^6 - R^4 + 1)\sin^2\omega + (R^2-1)^3\, (R^2+1)\sin^4\omega=0\,.
\label{VR2}
\ee

 In the undeformed theory, with $\omega=0$, only the first possibility
(\ref{VR1}) gives a stationary point, which is well known, namely
\be
R= 5^{1/4}\,,\qquad \rho= {\rm arccosh}\,
  \big(\fft{3}{\sqrt5}\big)\,,\qquad x=0\,.
\label{old}
\ee
As $\omega$ increase above zero, two valid stationary points arise from
(\ref{VR1}), one of which is a continuous
deformation of (\ref{old}), and the other of
which is a new stationary point that is absent at $\omega=0$.  It is not
possible to give analytic expressions for the values of $R$ at the stationary
points for generic $\omega$, owing to the high degree of the polynomial.
However, it is fairly straightforward to see that for each of the two solutions
the value of $R$ at the stationary point increases monotonically as $\omega$
increases through its range $\omega=0$ to $\omega=\pi/4$, with
\bea
\hbox{Solution 1}:\qquad && 0\le R\le \fft{\sqrt5 -1}{2}\,,\qquad \hbox{for}\
  0\le\omega\le\fft{\pi}{4}\,,\nn\\
\hbox{Solution 2}:\qquad && 5^{1/4} \le R \le \fft{\sqrt5 + 1}{2}\,,\qquad \hbox{for}\ 0\le\omega\le\fft{\pi}{4}\,.
\label{Sol12}
\eea

   The alternative factor (\ref{VR2}) in the stationarity condition
$\del V/\del R=0$ gives rise to just one branch of valid solutions for $R$.
Again, $R$ at the stationary point turns out to be monotonically increasing
as $\omega$ increases from 0 to $\pi/4$, with
\be
\hbox{Solution 3}:\qquad 0\le R\le 1\,,\qquad \hbox{for}\
0\le\omega\le\fft{\pi}{4}\,.\label{Sol3}
\ee
For all three of the solutions in (\ref{Sol12}) and (\ref{Sol3}) the value
of $\rho$ at the stationary point is given by (\ref{rhosol}), and they all have
$x=0$.

It is a simple matter to calculate the masses of the scalar fluctuations
around the various AdS$_4$ vacua.  Since the scalar $\chi$ does not appear
in the potential it is massless, and we shall not include it in the subsequent
discussion.  For the remaining scalars it is useful first to define a
rescaled field in place of $x$, so that all three of the scalars have the
same canonically-normalised kinetic terms.  Thus we may define
\be
\phi_1=\sigma\,,\qquad \phi_2=\rho\,,\qquad \phi_3=\sqrt3\, x\,.
\ee
The relevant parts of the Lagrangian for our present discussion then give
\be
{\cal L}=\sqrt{-g}\, \Big( R -
              \ft12\sum_{i=1}^3 (\del\phi_i)^2 -V(\phi)\Big)\,.
\ee
We may determine the masses of the scalar fluctuations, by
calculating the eigenvalues of the Hessian
matrix of second derivatives of $V$, evaluated at
the chosen stationary point.  Since the value of the potential at
the stationary
point, and hence the cosmological constant $\Lambda$, depends on the choice
of solution in (\ref{Sol12}) or (\ref{Sol3}),
and also on the value of the deformation parameter $\omega$, it is
advantageous to rescale the Hessian appropriately.  Since
$\Lambda=\ft12 V(\bar\phi)$, where $V(\bar\phi)$ denotes the value of
the scalar
potential at the stationary point $\phi_i=\bar\phi_i$, it is convenient to
calculate the rescaled Hessian matrix
\be
M_{ij} = \fft{8}{3 V(\bar\phi)}\,
    \fft{\del^2 V}{\del\phi_i\del\phi_j}\Big|_{\phi_k=\bar\phi_k}\,.
\ee
The eigenvalues of this matrix will give the three scalar masses normalised
by the mass $m_{BF}$ of the Breitenl\"ohner-Freedman bound, which is
given by
\be
m^2_{BF}= \ft34\, \Lambda\,.
\ee
Thus eigenvalues of $M_{ij}$ that are greater than or equal to $-1$ obey
the Breitenl\"ohner-Freedman bound.

   The usual AdS$_4$ solution gives an $M_{ij}$ that is already diagonal, with
\be
M_{ij}=\hbox{diag}\, \left( -\fft{8}{9},\, -\fft{8}{9},\, -\fft{8}{9}\right)\,,
\ee
and so, as is well known, the scalar masses are all equal and above the
BF bound.

   For Solution 1 and Solution 2, arising from the real positive roots
of (\ref{VR1}), we find
\be
\begin{pmatrix} \fft{8}{15} & \pm \fft{16}{5\sqrt3} & 0\cr
           \pm\fft{16}{5\sqrt3} & \fft{16}{15} & 0\cr
    0&0& -\fft{16}{15} \end{pmatrix}\,,
\ee
where the plus signs arise for Solution 1 and the minus signs for Solution 2.
Remarkably, the matrix is independent of the value of the parameter $\omega$.
The upper left $2\times2$ sub-matrix must be diagonalised to obtain the masses.
Upon doing this, we find
\be
\hbox{mass}^2 = \left( \fft{8}{3},\, -\fft{16}{15},\, -\fft{16}{15}\right)\,.
\ee
Thus two of the scalars have masses that violate the BF bound.
Although the  cosmological constant of the AdS$_4$ solution depends
on the value of $\omega$, the masses of the fluctuations, normalised
with respect to the cosmological constant, do not.  A similar phenomenon
was encountered in \cite{bodiguva} for AdS$_4$ vacua in the $SU(3)$-invariant
sector of the ${\cal N}=8$ theory.

  For Solution 3 we find that the matrix $M_{ij}$ is already diagonal,
with
\be
M_{ij}= \hbox{diag}\, \left(\fft{8}{3},\, \fft{8}{3},\, -\fft{4}{3}\right)\,.
\ee
Thus the scalar field $x$ violates the BF bound, while $\rho$ and $\sigma$
have positive mass-squared in this AdS$_4$ vacuum.  Again, the normalised
masses are independent of the value of the deformation parameter $\omega$.
If the scalar field $x$ were (consistently) truncated from the theory,
this AdS$_4$ vacuum would then be stable.



\begin{thebibliography}{99}

\bibitem{Dall'Agata:2012bb}
  G.~Dall'Agata, G.~Inverso and M.~Trigiante,
{\it Evidence for a family of $SO(8)$ gauged supergravity theories,}
  Phys.\ Rev.\ Lett.\  {\bf 109}, 201301 (2012),
 arXiv:1209.0760 [hep-th].

\bibitem{deWit:2013ija}
  B.~de Wit and H.~Nicolai,
{\it Deformations of gauged $SO(8)$ supergravity and supergravity in
  eleven dimensions},
  JHEP {\bf 1305}, 077 (2013),
  arXiv:1302.6219 [hep-th].

\bibitem{fispilwar}
  T.~Fischbacher, K.~Pilch and N.~P.~Warner,
{\it New supersymmetric and stable, non-supersymmetric phases in
supergravity and holographic field theory,}
  arXiv:1010.4910 [hep-th].

\bibitem{dewitnic} B. de Wit and H. Nicolai,
{\it $N=8$ supergravity},
Nucl.\ Phys.\ B {\bf 208}, 323 (1982).

\bibitem{Gauntlett:2009bh}
  J.~P.~Gauntlett, J.~Sonner and T.~Wiseman,
 {\it Quantum Criticality and Holographic Superconductors in M-theory,}
  JHEP {\bf 1002}, 060 (2010), arXiv:0912.0512 [hep-th].

\bibitem{Charmousis:2010zz} 
  C.~Charmousis, B.~Gouteraux, B.~S.~Kim, E.~Kiritsis and R.~Meyer,
{\it Effective Holographic Theories for low-temperature condensed matter systems,}
  JHEP {\bf 1011}, 151 (2010), arXiv:1005.4690 [hep-th].

\bibitem{Gouteraux:2011ce} 
  B.~Gouteraux and E.~Kiritsis,
{\it Generalized Holographic Quantum Criticality at Finite Density,}
  JHEP {\bf 1112}, 036 (2011), arXiv:1107.2116 [hep-th].
  
\bibitem{Bueno:2012vx} 
  P.~Bueno, W.~Chemissany and C.~S.~Shahbazi,
{\it On $hvLif$-like solutions in gauged Supergravity,}
  Eur.\ Phys.\ J.\ C {\bf 74}, no. 1, 2684 (2014), arXiv:1212.4826 [hep-th].
  
\bibitem{Gouteraux:2012yr} 
  B.~Gouteraux and E.~Kiritsis,
{\it Quantum critical lines in holographic phases with (un)broken symmetry,}
  JHEP {\bf 1304}, 053 (2013), arXiv:1212.2625 [hep-th].
  

\bibitem{Lifschytz:2009sz}
  G.~Lifschytz and M.~Lippert,
 {\it Holographic magnetic phase transition,}
  Phys.\ Rev.\ D {\bf 80}, 066007 (2009),
 arXiv:0906.3892 [hep-th].

\bibitem{D'Hoker:2010rz}
  E.~D'Hoker and P.~Kraus,
{\it Holographic Metamagnetism, Quantum Criticality, and Crossover Behavior,}
  JHEP {\bf 1005}, 083 (2010),
 arXiv:1003.1302 [hep-th].

\bibitem{Bergman:2012na}
  O.~Bergman, J.~Erdmenger and G.~Lifschytz,
{\it A Review of Magnetic Phenomena in Probe-Brane Holographic Matter,}
  Lect.\ Notes Phys.\  {\bf 871}, 591 (2013), arXiv:1207.5953 [hep-th].

\bibitem{Donos:2012yu}
  A.~Donos, J.~P.~Gauntlett, J.~Sonner and B.~Withers,
{\it Competing orders in M-theory: superfluids, stripes and metamagnetism,}
  JHEP {\bf 1303}, 108 (2013),
 arXiv:1212.0871 [hep-th].



\bibitem{Gubser:2008px}
  S.~S.~Gubser,
{\it Breaking an abelian gauge symmetry near a black hole horizon,}
  Phys.\ Rev.\ D {\bf 78}, 065034 (2008),
  arXiv:0801.2977 [hep-th].

\bibitem{Hartnoll:2008vx}
  S.~A.~Hartnoll, C.~P.~Herzog and G.~T.~Horowitz,
{\it Building a holographic superconductor,}
  Phys.\ Rev.\ Lett.\  {\bf 101}, 031601 (2008),
 arXiv:0803.3295 [hep-th].

\bibitem{Hartnoll:2008kx}
  S.~A.~Hartnoll, C.~P.~Herzog and G.~T.~Horowitz,
{\it Holographic superconductors,}
  JHEP {\bf 0812}, 015 (2008),
  arXiv:0810.1563 [hep-th].



\bibitem{Nakamura:2009tf}
  S.~Nakamura, H.~Ooguri and C.~S.~Park,
 {\it Gravity dual of spatially modulated phase,}
  Phys.\ Rev.\ D {\bf 81}, 044018 (2010),
 arXiv:0911.0679 [hep-th].

\bibitem{Ooguri:2010kt}
  H.~Ooguri and C.~S.~Park,
{\it Holographic end-point of spatially modulated phase transition,}
  Phys.\ Rev.\ D {\bf 82}, 126001 (2010),
arXiv:1007.3737 [hep-th].

\bibitem{Donos:2011bh}
  A.~Donos and J.~P.~Gauntlett,
{\it Holographic striped phases,}
  JHEP {\bf 1108}, 140 (2011),
arXiv:1106.2004 [hep-th].

\bibitem{Donos:2011qt}
  A.~Donos, J.~P.~Gauntlett and C.~Pantelidou,
{\it Spatially modulated instabilities of magnetic black branes,}
  JHEP {\bf 1201}, 061 (2012),
arXiv:1109.0471 [hep-th].

\bibitem{SCvsStripes}
J~Chang \emph{et al.},
{\it Direct observation of competition between superconductivity and charge
density wave order in $YBa_2Cu_3 O_{6.67}$,}
Nature Physics 8, 871–876 (2012).



\bibitem{Gubser:2009cg}
  S.~S.~Gubser and A.~Nellore,
{\it Ground states of holographic superconductors,}
  Phys.\ Rev.\ D {\bf 80}, 105007 (2009),
arXiv:0908.1972 [hep-th].

\bibitem{Horowitz:2009ij}
  G.~T.~Horowitz and M.~M.~Roberts,
{\it Zero temperature limit of holographic superconductors,}
  JHEP {\bf 0911}, 015 (2009),
 arXiv:0908.3677 [hep-th].

\bibitem{Bhattacharya:2014dea}
  J.~Bhattacharya, S.~Cremonini and B.~Gout\'eraux,
 {\it Intermediate scalings in holographic RG flows and conductivities,}
  arXiv:1409.4797 [hep-th].



\bibitem{Ito}
A. Ito \emph{et al.},
{\it Study of Ising system Fe$_x$Mn$_{1-x}$TiO$_3$ with exchange frustrations
by observing magnetization process},
J. of Magnetism and Magnetic Materials, {\bf 104-107}, 1635 (1992).

\bibitem{Kaczmarsca}
K. Kaczmarsca \emph{et al.},
{\it Magnetic, resistivity and ESR studies of the compounds
GdNi$_2$Sb$_2$ and GdCu$_2$Sb$_2$},
J. of Magnetism and Magnetic Materials, {\bf 147}, 81 (1995).

\bibitem{Grigera}
S. A. Grigera \emph{et al.}, {\it Magnetic field-tuned quantum criticality
in the metallic
ruthenate $Sr_3 Ru_2 O_7$,}
Science {\bf 294}, 329 (2001).

\bibitem{Krey}
C.~Krey \emph{et al.},
{\it First order metamagnetic transition in $H\!o_2Ti_2 O_7$
observed by vibrating coil magnetometry at milli-Kelvin temperatures,}
Phys.\ Rev.\ Lett.\ {\bf 108} 257204 (2012).

\bibitem{Lu:2013ura}
  H.~L\"u, Y.~Pang and C.~N.~Pope,
  {\it AdS dyonic black hole and its thermodynamics,}
  JHEP {\bf 1311}, 033 (2013), arXiv:1307.6243 [hep-th].

\bibitem{Lu:2014fpa}
  H.~L\"u, Y.~Pang and C.~N.~Pope,
  {\it An $\omega$ deformation of gauged STU supergravity,}
  JHEP {\bf 1404}, 175 (2014), arXiv:1402.1994 [hep-th].

 \bibitem{Hertog:2004dr}
  T.~Hertog and K.~Maeda,
  {\it Black holes with scalar hair and asymptotics in $N = 8$ supergravity,}
  JHEP {\bf 0407}, 051 (2004), hep-th/0404261.

  \bibitem{Ashtekar:1999jx}
  A.~Ashtekar and A.~Magnon,
 {\it Asymptotically anti-de Sitter space-times,}
 Class.\ Quant.\ Grav.\ {\bf 1}, L39 (1984).
  A.~Ashtekar and S.~Das,
 {\it Asymptotically anti-de Sitter space-times: Conserved quantities,}
  Class.\ Quant.\ Grav.\  {\bf 17}, L17 (2000),
 hep-th/9911230.


\bibitem{Cremonini:2012ir}
  S.~Cremonini and A.~Sinkovics,
 {\it Spatially modulated instabilities of geometries with hyperscaling
violation,}
  JHEP {\bf 1401}, 099 (2014),
arXiv:1212.4172 [hep-th].

\bibitem{Iizuka:2013ag}
  N.~Iizuka and K.~Maeda,
 {\it Stripe instabilities of geometries with hyperscaling violation},
  Phys.\ Rev.\ D {\bf 87}, no. 12, 126006 (2013),
arXiv:1301.5677 [hep-th].

\bibitem{Cremonini:2013epa}
  S.~Cremonini,
 {\it Spatially modulated instabilities for scaling solutions at finite
charge density},
  arXiv:1310.3279 [hep-th].

\bibitem{bodiguva} A.~Borghese, G.~Dibitetto, A.~Guarino, D.~Roest and
O.~Varela,
{\it The $SU(3)$-invariant sector of new maximal supergravity,}
  JHEP {\bf 1303}, 082 (2013),
  arXiv:1211.5335 [hep-th].


\end{thebibliography}
\end{document}